\documentclass[11pt]{article}

\usepackage[utf8]{inputenc}
\usepackage[T1]{fontenc}
\usepackage{amsmath,amssymb,amsfonts}
\usepackage{upgreek}
\usepackage{graphicx}
\usepackage{float}
\usepackage{microtype}
\usepackage{booktabs}
\usepackage{array}
\usepackage{multirow}
\usepackage{authblk}
\usepackage[hidelinks]{hyperref}
\usepackage{geometry}
\usepackage{caption}
\usepackage{times}
\usepackage{xcolor}

\title{\textbf{Retrieval-Augmented Generation-Based Color Restoration\\ for Low-Light Image Enhancement}}

\author{Li-Wei Lu}
\author{Shaou-Gang Miaou\thanks{Corresponding author: Shaou-Gang Miaou (e-mail: miaou@cycu.edu.tw). Shaou-Gang Miaou and Li-Wei Lu are with the Department of Electronic Engineering, Chung Yuan Christian University, Taoyuan 320, Taiwan.}}
\affil{Department of Electronic Engineering, Chung Yuan Christian University, Taoyuan 320, Taiwan}

\date{}

\begin{document}
\maketitle

\begin{abstract}
Recent low-light image enhancement (LLIE) methods have driven brightness and structural fidelity close to that of normally-exposed images, yet their outputs still exhibit systematic color shifts---skies drifting green, faces turning yellow, white objects reading warm. We attribute this to the fact that end-to-end LLIE training couples brightness, structure, and color within a single network, leaving the color channels weakly supervised. We recast color restoration as an independent sub-problem and decouple it from brightness enhancement, realizing it as a general-purpose post-processing module built on retrieval-augmented generation (RAG). Rather than relying solely on the parametric color priors learned during training, the module dynamically retrieves a reference image from an external high-quality color knowledge base and injects its color distribution into a color-restoration network to correct residual bias. The design comprises three components: (i) a dual-index FAISS retriever built on intermediate VGG19 features, capturing textural and structural similarity through global mean and variance statistics respectively; (ii) GlobalSPHistAdaIN, which reduces the reference image's spatial-preserving color histogram to a global color vector and modulates network features via adaptive instance normalization, removing any dependence on pixel-level spatial correspondence; and (iii) a residual formulation that predicts a color correction over the front-end output rather than regenerating the color channels. Across LOLv1, LOLv2-Real, and LOLv2-Synthetic, the module consistently improves color-specific metrics: with CPGA-Net++ as the front end, it lowers $\Delta E_{2000}$ from 8.91 to 8.40 and $\mathrm{MAE}_{ab}$ from 5.27 to 4.71 on LOLv1, with color gains markedly exceeding brightness gains. The module remains effective when the front end is swapped for LLFormer, FLIGHTNet, or IAT, confirming cross-front-end generality. Ablations isolate the contribution of each component and, notably, show that a VGG19 dual index outperforms CLIP-based retrieval---indicating that color restoration depends on textural and structural similarity rather than high-level semantics---while substituting the training-time knowledge base (DIV2K) with an unseen one (Flickr2K) retains most of the gain, confirming that the network learns to exploit retrieved color statistics rather than memorizing a fixed prior.

\vspace{0.5em}
\end{abstract}

\section{Introduction}

Human vision relies on adequate illumination to perceive the color, texture, and spatial structure of a scene. Many practical settings, however, cannot guarantee such conditions---nighttime surveillance, dashcam recording, indoor photography without flash, medical endoscopy, and the night-vision systems of autonomous vehicles all routinely produce low-light images marked by low contrast, elevated noise, lost detail, and distorted color. Beyond degrading the experience of a human observer, these degradations directly lower the accuracy of downstream vision tasks such as object detection, face recognition, and semantic segmentation. Recovering a clear and chromatically natural image from a low-light input has therefore been a long-standing problem in computer vision and image processing, studied under the name low-light image enhancement (LLIE).

Early LLIE methods were dominated by classical image-processing techniques, chiefly histogram equalization and Retinex theory~\cite{ref1}. The former redistributes pixel intensities to expand the dynamic range, while the latter models an image as the product of illumination and reflectance and recasts enhancement as an illumination-estimation problem, a line of thought that later gave rise to methods such as LIME~\cite{ref2}. Deep learning has since moved the field into a new phase: RetinexNet~\cite{ref3} realized the Retinex decomposition with neural networks and introduced the first large-scale paired LOL dataset~\cite{ref3}, and subsequent models---LLFormer~\cite{ref4}, CPGA-Net~\cite{ref5} and its extension CPGA-Net++~\cite{ref6}, FLIGHTNet~\cite{ref7}, and IAT~\cite{ref8}---have steadily pushed PSNR and SSIM on the LOL benchmarks to the point where brightness and detail recovery approach that of normally-exposed images.

Progress in brightness recovery, however, has not translated into comparable progress in color recovery. Although the outputs of current LLIE models are competitive in luminance, they commonly display systematic color shifts---greenish skies, yellowish skin tones, warm-tinted white objects. The root of this behavior is that end-to-end training tends to couple brightness, structure, and color within a single network, where the correction of chromatic channels is easily dominated by the luminance- and structure-reconstruction objectives. Compounding this, most LLIE losses optimize overall pixel reconstruction error or perceptual similarity; in RGB, pixel losses are typically more sensitive to luminance than to chromatic error, and perceptual losses reward texture and structure rather than chromatic accuracy. Subtle color deviations are thus weakly constrained and remain a secondary concern throughout training. At inference, moreover, a model predicts largely from the parametric color priors absorbed during training, which cannot fully cover the diverse color distributions of real low-light scenes; inputs that fall outside the training distribution or involve complex illumination may yield results that are numerically plausible yet chromatically inconsistent with the true scene.

This bottleneck echoes a challenge already addressed in natural language processing, where retrieval-augmented generation (RAG)~\cite{ref9} couples a parametric generator with an external non-parametric knowledge base, letting the model retrieve query-relevant information at inference time to supplement what its weights encode. Transposed to low-light color restoration, the residual color bias can be viewed as the visual uncertainty a model exhibits when it lacks an external chromatic reference, and an external repository of high-quality color images can play the role of the knowledge base. By dynamically retrieving a reference image similar to the input scene and injecting its color distribution as a conditional prior, the model no longer depends solely on its internal statistics but instead grounds its correction in real reference imagery.

Building on this idea, we decouple color restoration from the front-end LLIE model and cast it as a general-purpose post-processing module that operates on top of a frozen enhancement network. Because the front end is never retrained, the module applies a consistent chromatic correction across different front ends. Three observations motivate its design. First, existing evaluations under-report color: PSNR, SSIM, and LPIPS~\cite{ref10} remain the default metrics, and although LPIPS~\cite{ref10} reflects perceptual similarity it is grounded in deep features attuned to structure and texture, leaving pure chromatic deviation relatively invisible---so color restoration deserves its own metrics and a dedicated design. Second, adding an external color prior without disturbing the recovered brightness is non-trivial: reference-guided colorization offers a template, but its methods assume grayscale inputs and pixel-level correspondence, whereas an LLIE front end already supplies an initial color estimate and a retrieved reference may differ substantially from the input in scene layout, making spatially-aligned injection unreliable. Third, RAG~\cite{ref9} has been extensively studied in language and multimodal tasks but rarely instantiated for low-level image restoration, so its use as a retrievable color prior for LLIE lacks concrete empirical validation.

We address these points with a module comprising a dual-index FAISS~\cite{ref11} retriever over intermediate VGG19~\cite{ref12} features, a global color-injection mechanism (GlobalSPHistAdaIN) that dispenses with spatial correspondence, and a residual learning formulation that focuses the network on correcting rather than regenerating color. Our contributions are as follows:

\begin{itemize}
\item We instantiate RAG~\cite{ref9} for the color-restoration sub-problem of LLIE, establishing a complete and reproducible pipeline---external color knowledge base $\rightarrow$ dual-index FAISS~\cite{ref11} retrieval $\rightarrow$ neural color injection---and thereby introducing the notion of supplementing color priors from external knowledge into low-light enhancement.
\item We propose GlobalSPHistAdaIN, which pairs a globally-pooled SPHist color vector with AdaIN~\cite{ref13}-based modulation. Removing spatial correspondence lifts the computational ceiling on high-resolution inputs, avoids mis-injected color when the reference and input scenes diverge, and---together with residual learning---improves training stability and cross-dataset generalization.
\item We systematically compare retrieval strategies and find that a VGG19~\cite{ref12} dual index outperforms CLIP~\cite{ref14}-based semantic retrieval, offering evidence that color restoration relies on textural and structural similarity rather than high-level semantic similarity---an observation of practical value for the design of visual retrieval systems.
\item We evaluate the module quantitatively on paired datasets and via no-reference metrics on unpaired data, showing stable improvement in color-specific metrics over front-end baselines and a consistent trend across different front ends, which demonstrates the module's cross-model generality rather than a fit to one architecture.
\end{itemize}

\section{Related Work}

\subsection{Low-Light Image Enhancement}

Classical LLIE methods fall into two main lines. Histogram equalization improves contrast by remapping an image's intensity distribution toward uniformity, but global equalization tends to over-enhance already-balanced regions and cannot adapt to spatially varying illumination; adaptive variants and contrast-limited adaptive histogram equalization (CLAHE)~\cite{ref15} mitigate this by equalizing local tiles and capping local histogram amplification to suppress noise. The Retinex line~\cite{ref1} instead models an observed image as the product of illumination and reflectance, turning enhancement into illumination estimation; multi-scale Retinex with color restoration (MSRCR)~\cite{ref16} adds a color-compensation term to counter the grayness and chromatic distortion that multi-scale enhancement can introduce, and LIME~\cite{ref2} initializes the illumination map from the per-pixel maximum across RGB channels and refines it with a structure prior, achieving good visual quality at low cost. These methods are interpretable, data-free, and inexpensive, but their reliance on hand-crafted priors and parameters makes it hard to jointly recover brightness, detail, and color under severe degradation.

Deep learning has substantially advanced the field. RetinexNet~\cite{ref3} jointly trains an illumination-adjustment and a decomposition network end-to-end and contributes the first large-scale paired LOL dataset~\cite{ref3}, laying the groundwork for supervised approaches. LLFormer~\cite{ref4} employs a Transformer to model long-range dependencies and reports competitive results across several benchmarks. More recent work emphasizes physical priors alongside efficiency: IAT~\cite{ref8} remains competitive at very low parameter counts and suits real-time use. CPGA-Net~\cite{ref5} combines a dark channel prior, a bright channel prior, YCbCr luminance, and gamma correction, integrating these cues through an intersection-aware adaptive fusion module to approach state-of-the-art quality at roughly 0.025~M parameters. Two extensions follow on the same architecture: CPGA-Net+ reintroduces atmospheric-scattering priors into the attention design and localizes gamma correction, while CPGA-Net++~\cite{ref6} further exploits the local-processing branch with an improved fusion module and ConvNeXt blocks, attaining strong image quality across benchmarks. FLIGHTNet~\cite{ref7} adopts a lightweight design that improves restoration through progressive feature fusion.

Despite this progress on luminance- and structure-related metrics, color-specific metrics rarely appear in mainstream evaluation---a gap that reflects a structural limitation of end-to-end training on the color channels. Most LLIE models are supervised primarily by L1 or L2 pixel losses with a perceptual loss as auxiliary; yet in RGB, pixel losses are generally more sensitive to luminance than to chromatic error, and perceptual losses are grounded in texture and structure rather than explicit color supervision. As a result, a model may recover luminance and texture well while still producing an overall color cast or local chromatic drift---distortions that PSNR, SSIM, and LPIPS~\cite{ref10} do not fully surface, since even LPIPS~\cite{ref10} is not designed to isolate chromatic accuracy and its color-sensitive component is easily diluted when brightness and structure improve markedly. Because low-light color information is itself compromised by noise and reduced saturation, end-to-end training alone struggles to reliably supply the missing chromatic cues; we therefore treat color restoration as a separate stage and provide an additional color prior from external high-quality imagery.

\subsection{Color Restoration and Reference-Based Colorization}

The design of a color-restoration network, its losses, and its metrics all depend on a color space that expresses chromatic difference reliably. RGB corresponds directly to sensor output but couples its three channels tightly, and Euclidean distance in RGB does not correspond linearly to perceived color difference. The CIE Lab space~\cite{ref17}, defined by the CIE in 1976, decomposes color into a luminance channel $L$ and two chromatic channels $a$ (green--red) and $b$ (blue--yellow), and offers two properties central to this work: perceptual uniformity---Euclidean distance in Lab aligns well with perceived difference, which underlies measures such as $\Delta E_{2000}$---and a clear separation between luminance and chroma, so that adjusting $a$ and $b$ generally leaves the $L$-encoded brightness structure intact. When the front end has already restored luminance to near-normal levels, the color-restoration network can operate in Lab with $L$ as a brightness condition and $a$, $b$ as prediction targets, keeping luminance error from contaminating the color supervision. We accordingly define our color-injection mechanism, color losses, and color-specific metrics ($\Delta E_{2000}$, $\mathrm{MAE}_{ab}$) in Lab.

Automatic colorization predicts a full-color image from a grayscale input with no external reference. Iizuka et al.~\cite{ref18} jointly learn global and local priors, combining semantic classification with pixel-level color mapping, while CIC~\cite{ref19} casts colorization as classification over quantized color bins and rebalances classes to counter skewed color distributions. The core limitation is inherent ambiguity: a single grayscale input admits many plausible colorings, and because models learn the high-probability average of the training distribution, outputs tend toward desaturated, conservative colors that lack diversity.

Reference-based colorization instead supplies a color-similar reference image as a prior, converting an ill-posed generation problem into a color-transfer problem. Early methods matched intensity and texture features to transfer reference color~\cite{ref20,ref21}, and Gupta et al.~\cite{ref22} retrieved similar images from a web gallery. In the deep-learning era, deep exemplar-based colorization~\cite{ref23} uses a similarity network built on intermediate VGG19~\cite{ref12} features to establish cross-image correspondence and a colorization network to transfer color accordingly. Pik-Fix~\cite{ref24} is the most direct inspiration for our design: it models old-photo restoration as three cooperating sub-tasks---a restoration network based on a multi-level Residual Dense Network (multi-level RDN) that repairs degradation and structural damage, a similarity network that uses pretrained ResNet34~\cite{ref25} features and a Similarity Sub-Net to build the input--reference spatial correspondence and spatially align the reference features, and a colorization network with a global U-Net~\cite{ref26} backbone into which the spatially-aligned color features are injected at each decoder layer. Pik-Fix~\cite{ref24} also introduces a VGG19~\cite{ref12}-based reference-retrieval scheme that measures textural similarity by the feature global mean and structural similarity by variance and covariance, then selects the most similar reference from the training data by a weighted combination, showing stable color restoration and a degree of robustness to reference-selection error. Our retrieval design adopts this idea of describing an image by the mean and variance statistics of its VGG19~\cite{ref12} features, and extends it into a dual-index FAISS~\cite{ref11} retriever so that texture and structure similarity are indexed separately and combined at query time.

Color histograms have long served image retrieval and color transfer~\cite{ref27}, being robust to spatial shifts, viewpoint, and partial occlusion; but a conventional global histogram discards spatial layout---two images with swapped sky and ground regions may share nearly identical global histograms while corresponding to entirely different spatial color arrangements---and its hard binning is non-differentiable, hindering integration into end-to-end frameworks. To overcome both issues, Pik-Fix~\cite{ref24} introduces the spatial-preserving color histogram (SPHist), which replaces hard assignment with a differentiable soft assignment, distributing each pixel probabilistically across color bins while retaining spatial dimensions; SPHist thus preserves color-distribution characteristics, supports back-propagation, and serves well as a color intermediary in deep frameworks.

In style transfer and conditional generation, a recurring question is how to inject the global style of one image into another at low cost. Instance normalization (IN)~\cite{ref28} normalizes each channel of a single image by its own mean and standard deviation, reducing the influence of the image's own style statistics, but it only standardizes features without introducing new style information and generalizes poorly across styles. Adaptive instance normalization (AdaIN)~\cite{ref13} addresses this by modulating content features with the statistics of a style feature---first standardizing with the content's own mean and standard deviation, then rescaling to the style's statistics---so that style injection completes in a single forward pass without per-style training. Because its modulation parameters can be computed on the fly from any input, AdaIN~\cite{ref13} generalizes flexibly and has been extended to conditional generation, image translation, and feature modulation. We build on this property, treating AdaIN~\cite{ref13} as a means of modulating intermediate features with external image statistics and applying it to inject an external color prior for low-light color restoration.

\subsection{Retrieval-Augmented Generation and Its Extension to Vision}

RAG~\cite{ref9}, introduced by Lewis et al., integrates the parametric knowledge of a large language model with the non-parametric knowledge of an external base. A typical system pairs a retriever, which finds relevant fragments in the external base given a query, with a generator, which conditions on both query and retrieved content to produce output; this mitigates two limitations of purely parametric models---the high cost of updating knowledge and the lack of external grounding.

Efficient similarity search over large high-dimensional collections is central to retrieval. FAISS~\cite{ref11} provides approximate-nearest-neighbor search with several index structures---flat inner-product (IndexFlatIP), inverted file, hierarchical navigable small-world graphs, and product quantization---that trade off speed, memory, and accuracy. When vectors are L2-normalized, the inner product computed by IndexFlatIP equals cosine similarity, which makes it a common choice for image and text feature retrieval; we adopt FAISS~\cite{ref11} as the basis of our retrieval pipeline.

The choice of feature extractor strongly shapes retrieval behavior in vision. VGG19~\cite{ref12}, trained on ImageNet classification, yields intermediate feature maps that effectively capture textural and structural statistics; Zhang et al.~\cite{ref10} note that deep convolutional features align well with human perceptual judgments, with VGG features particularly reflecting texture and structure differences, and Pik-Fix~\cite{ref24} uses VGG19~\cite{ref12} feature means for texture and variance/covariance for structure---a design that informs our dual index. CLIP~\cite{ref14}, trained by contrastive learning on large image--text pairs, extracts features highly aligned with textual semantics, encoding concepts such as people, clothing, and color into a shared multimodal space, which suits cross-modal retrieval and semantic classification. The essential difference lies in how each defines similarity: VGG19~\cite{ref12} leans toward textural and structural similarity, whereas CLIP~\cite{ref14} emphasizes cross-modal semantic similarity. For color restoration, an ideal reference should share texture, structure, and color distribution with the target, so an effective color prior likely depends more on low- to mid-level visual similarity than on semantic agreement---two semantically similar images may still differ markedly in color.

Although RAG~\cite{ref9} is well studied in language, its extension to vision remains early. Multimodal RAG~\cite{ref29} focuses on cross-modal retrieval and joint generation for tasks requiring semantic reasoning, such as visual question answering and image captioning. In generation, retrieval-augmented diffusion models~\cite{ref30} introduce retrieved similar images as conditioning to improve diversity and consistency. Yet the specific problem of supplementing a color prior from an external image base to improve restoration quality lacks systematic empirical study. We extend RAG~\cite{ref9} to low-light color restoration, examine the feasibility of an external image knowledge base for color-prior supplementation, and analyze how retrieval quality affects the final result---shifting the focus from semantic reasoning and generation toward external color priors in a visual restoration task.

\subsection{Research Gaps}

Three gaps emerge from this review. First, existing LLIE methods bind brightness enhancement and color restoration in a single end-to-end pipeline, leaving color supervision comparatively weak and color-accuracy gains less pronounced than luminance gains; improving color accuracy independently, without disturbing existing brightness gains, is a key opening. Second, reference-guided colorization, though mature, is premised on grayscale inputs and physical-damage repair, which differs fundamentally from the low-light setting and leads to input-condition mismatch and unreliable cross-dataset spatial correspondence when applied directly. Third, RAG~\cite{ref9} is well established in language but has scarce concrete validation in visual color restoration; realizing external-base retrieval with conditional generation as a working LLIE color-restoration system and validating each design choice offers a reproducible application case for RAG~\cite{ref9} in vision. These observations frame the retrieval-based color-restoration framework developed in the following section.

\section{Method}

\subsection{Overview}

\begin{figure}[htbp]
\centering
\includegraphics[width=0.62\columnwidth]{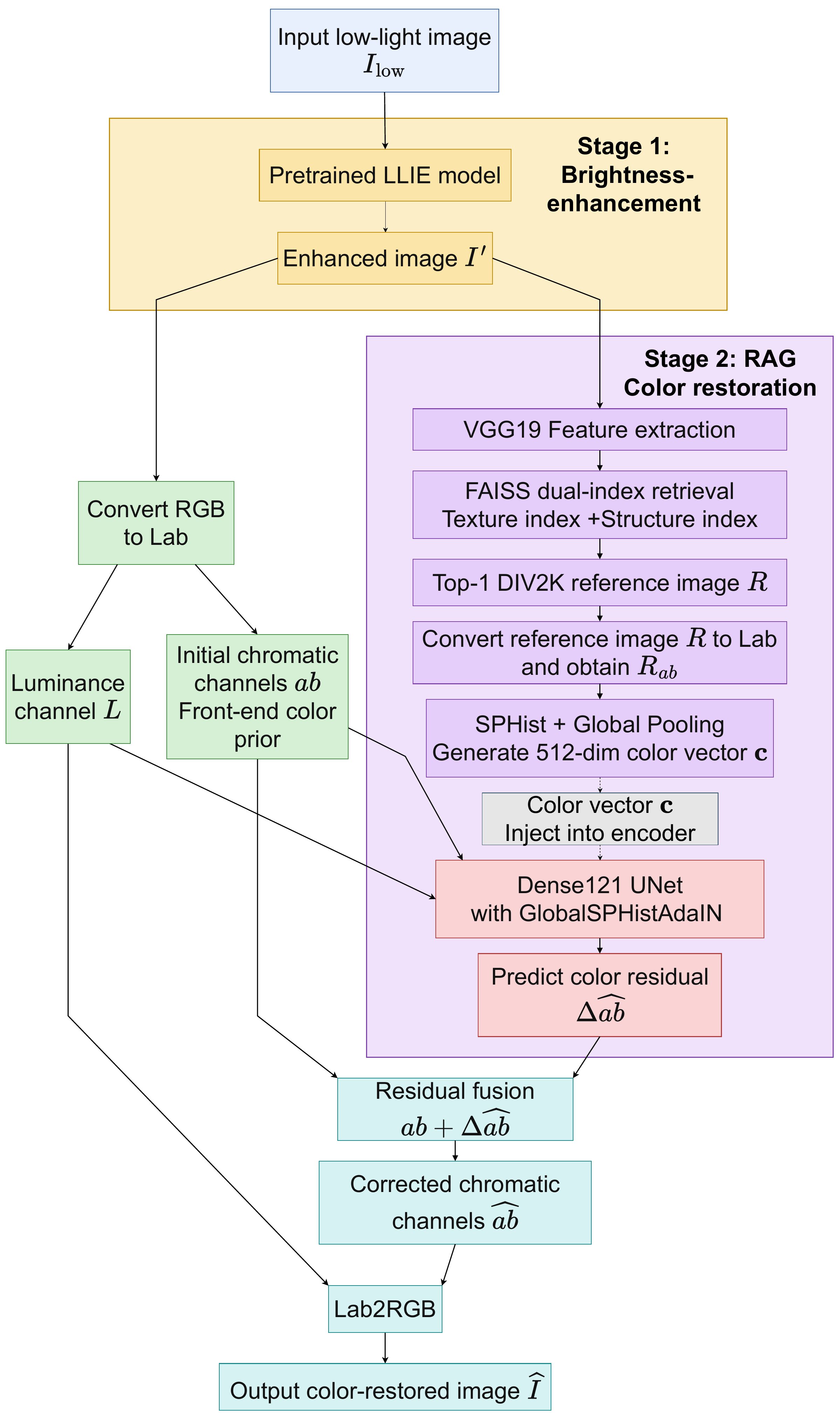}
\caption{Overall system architecture. Stage 1 (frozen LLIE front end) produces a brightness-enhanced image; Stage 2 retrieves a reference from an external knowledge base and injects its color distribution to predict a color residual.}
\label{fig:overview}
\end{figure}

We propose a two-stage system for low-light color restoration (Fig.~\ref{fig:overview}). The first stage uses an existing LLIE model as a brightness-enhancement front end; the second stage is a color-restoration module, designed around the RAG~\cite{ref9} concept, that uses the color distribution of an external reference image to correct the residual color bias and distortion left in the enhanced output.

\textit{Stage 1 --- Brightness enhancement.} Given a raw low-light image $I_{\mathrm{low}} \in \mathbb{R}^{H \times W \times 3}$, a pretrained front end produces a brightness-enhanced image $I'$. We evaluate four front-end configurations---CPGA-Net++~\cite{ref6}, LLFormer~\cite{ref4}, FLIGHTNet~\cite{ref7}, and IAT~\cite{ref8}---each representing a distinct architectural philosophy. Throughout second-stage training the front end stays frozen and receives no gradient updates; its output serves as the fixed input to the color-restoration module. We convert $I'$ from RGB to Lab and split it into a luminance channel $L \in \mathbb{R}^{H \times W \times 1}$ and chromatic channels $ab \in \mathbb{R}^{H \times W \times 2}$. The $L$ channel is the brightness reference for all downstream modules, and $ab$ retains the front end's initial color estimate as a local color prior.

\textit{Stage 2 --- RAG color restoration.} The second stage comprises three sub-modules: reference retrieval, color-vector extraction, and color-residual prediction. A FAISS~\cite{ref11} retriever uses deep features of the enhanced image to retrieve the visually nearest reference $R$ from a prebuilt DIV2K~\cite{ref31} index, converts it to Lab, and takes its chromatic channels $R_{ab}$ as the color source. The retriever uses a dual-index design: global mean and global variance vectors of VGG19~\cite{ref12} features describe textural and structural characteristics respectively, and a weighted re-ranking selects the reference best suited to serve as a color prior. From $R_{ab}$, the color-vector extraction module computes a spatial-preserving color histogram (SPHist) and reduces it, by global average pooling, to a 512-dimensional color vector $\mathbf{c}$. Finally, the color-restoration network takes $\mathrm{cat}(L, ab)$ as its backbone input and injects $\mathbf{c}$ into encoder features through the proposed GlobalSPHistAdaIN module. Rather than predicting the full chromatic channels, the network predicts a color residual $\Delta\widehat{ab}$ that is added back to the front end's initial $ab$. The first stage thus improves brightness and visibility while the second focuses on correcting residual color bias.

\subsection{Brightness-Enhancement Front End}

Applying color restoration directly to a raw low-light image forces a single network to contend simultaneously with luminance degradation and color distortion, increasing learning difficulty. We therefore place an existing LLIE model first to produce a brightness-stabilized image, and let the second-stage RAG~\cite{ref9} module handle color correction---splitting the overall task into a brightness sub-problem and a color sub-problem so that the second stage can concentrate on residual color bias. CPGA-Net++~\cite{ref6} is our primary front end, chosen for its lightweight, theory-driven design and stable enhancement; LLFormer~\cite{ref4}, FLIGHTNet~\cite{ref7}, and IAT~\cite{ref8} are used in a subset of experiments not to re-compare the front ends themselves but to supply differently-styled enhanced outputs, testing whether the color module retains its effect across front ends. Since these models differ in brightness lift, color preservation, noise suppression, and exposure correction, consistent color improvement across their outputs would indicate that the method is not tied to CPGA-Net++~\cite{ref6} and possesses genuine modular, cross-model applicability.

In implementation, each front end acts solely as a Stage-1 enhancer. The enhanced $I'$ is converted to Lab; the $L$ channel becomes the luminance/structure reference for Stage 2, and the $ab$ channels provide the front end's initial color estimate. During data loading, training and validation images are cropped to multiples of 32 to satisfy the down-/up-sampling requirements of the Dense121 U-Net; at inference, images whose dimensions are not multiples of 32 are padded to the nearest multiple by reflect padding and cropped back to their original size afterward, so that padded regions do not affect the output. We normalize the channels to $[-1, 1]$: $L$ from $[0, 100]$ by dividing by 50 and subtracting 1, and $ab$ from $[-110, 110]$ by dividing by 110. The color-restoration network outputs a residual $\Delta\widehat{ab}$, and the corrected chromatic channels and final image are
\begin{equation}
\widehat{ab} = ab + \Delta\widehat{ab}
\end{equation}
\begin{equation}
\hat{I} = \mathrm{Lab2RGB}(L, \widehat{ab})
\end{equation}

\subsection{FAISS-Based Reference Retrieval}
\label{sec:retrieval}

\textit{VGG19 feature extraction.} We use an ImageNet-pretrained VGG19~\cite{ref12} as the feature backbone for retrieval, with weights frozen, and take the output of the ReLU activation in the third convolutional block (256 channels) as the image representation. Relative to shallower features it offers more stable texture description, and relative to deeper features it is less biased toward object semantics, making it well suited to measuring cross-scene textural and structural similarity. Inputs are resized to $256 \times 256$ and normalized by the ImageNet mean and standard deviation; the extracted feature map is
\begin{equation}
\mathbf{F}_i \in \mathbb{R}^{C \times H' \times W'}, \quad C = 256
\end{equation}
where $i$ indexes the image and $H', W'$ are the spatial dimensions of the feature map. From it we compute a global mean vector $\mathbf{m}_i$ and a global variance vector $\mathbf{v}_i$:
\begin{equation}
\mathbf{m}_i = \frac{1}{H'W'} \sum_{h,w} \mathbf{F}_i(\cdot, h, w), \quad \mathbf{m}_i \in \mathbb{R}^{C}
\end{equation}
\begin{equation}
\mathbf{v}_i = \frac{1}{H'W'} \sum_{h,w} \big(\mathbf{F}_i(\cdot, h, w) - \mathbf{m}_i\big)^2, \quad \mathbf{v}_i \in \mathbb{R}^{C}
\end{equation}
where $(\cdot)$ denotes all elements along the channel dimension. The global mean $\mathbf{m}_i$ corresponds to the image's overall textural style and the global variance $\mathbf{v}_i$ to local structure and contrast variation. Both are L2-normalized so that a subsequent inner product equals cosine similarity:
\begin{equation}
\hat{\mathbf{m}}_i = \frac{\mathbf{m}_i}{\lVert \mathbf{m}_i \rVert_2}, \quad \hat{\mathbf{v}}_i = \frac{\mathbf{v}_i}{\lVert \mathbf{v}_i \rVert_2}
\end{equation}

\begin{figure}[htbp]
\centering
\includegraphics[width=0.7\columnwidth]{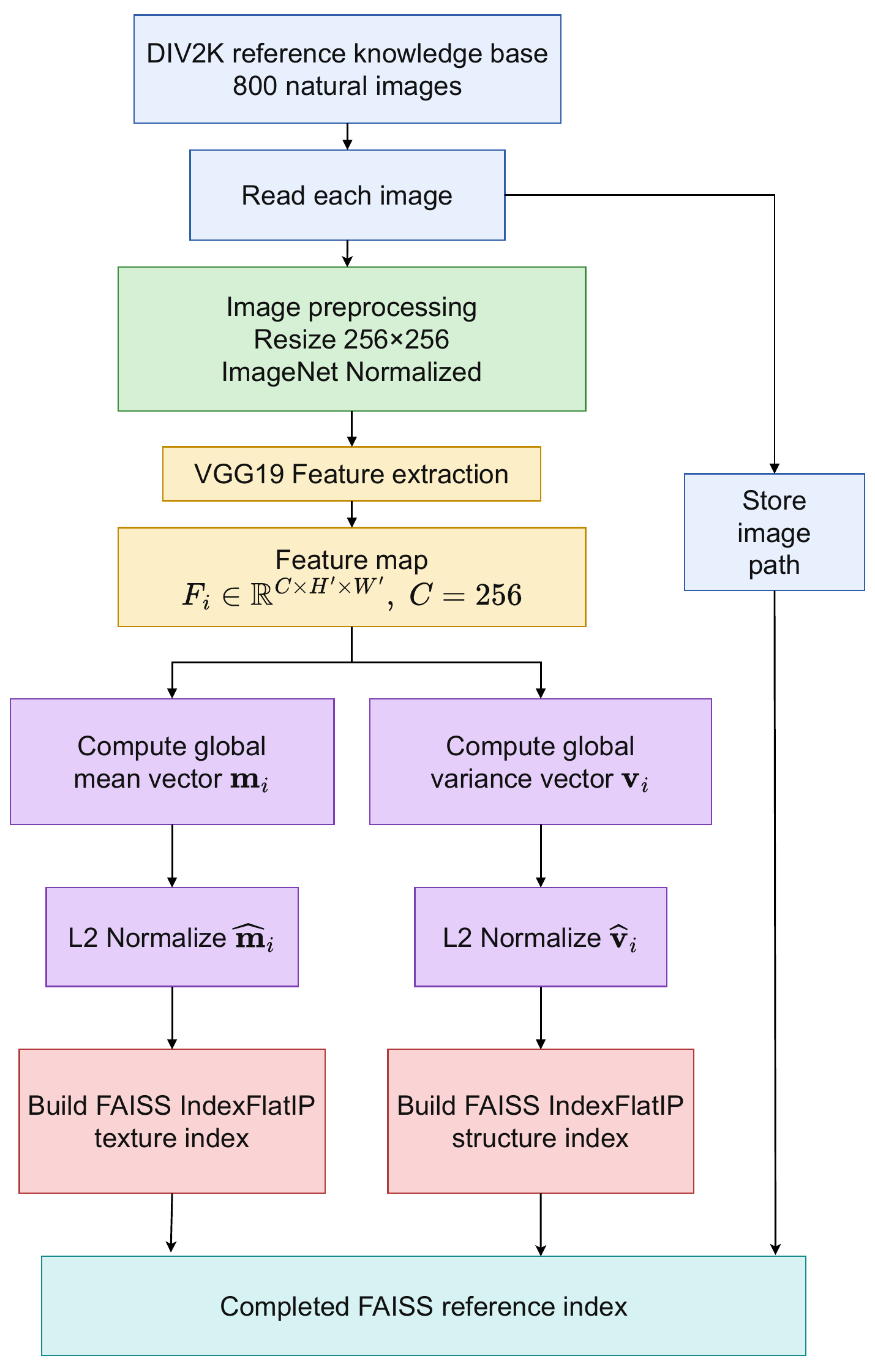}
\caption{Dual-index construction: VGG19 features of each knowledge-base image are reduced to global mean and variance vectors, stored in a texture index and a structure index respectively.}
\label{fig:dualindex}
\end{figure}

\textit{Dual-index design.} We use DIV2K~\cite{ref31} as the reference knowledge base for its diverse natural imagery. During index construction, VGG19~\cite{ref12} features are extracted for every DIV2K~\cite{ref31} image, and $\hat{\mathbf{m}}_i$ and $\hat{\mathbf{v}}_i$ are stored in two separate FAISS~\cite{ref11} indices: a texture index $\mathrm{Index}_{\mathbf{m}}$ holding the global mean vectors and a structure index $\mathrm{Index}_{\mathbf{v}}$ holding the global variance vectors (Fig.~\ref{fig:dualindex}). Because the DIV2K~\cite{ref31} base is modest in scale, we adopt the exact-search IndexFlatIP; as all vectors are L2-normalized beforehand, its inner product equals cosine similarity,
\begin{equation}
\mathrm{sim}(\mathbf{a}, \mathbf{b}) = \mathbf{a}^{\top}\mathbf{b} = \lVert \mathbf{a} \rVert \lVert \mathbf{b} \rVert \cos\theta = \cos\theta
\end{equation}
The dual-index design keeps texture and structure information independent. Concatenating the mean and variance vectors into a single index would fix their relative influence at concatenation time, whereas two indices can be queried separately and combined at re-ranking through weights $\alpha$ and $\beta$, giving the retrieval strategy greater flexibility.

\begin{figure}[htbp]
\centering
\includegraphics[width=0.6\columnwidth]{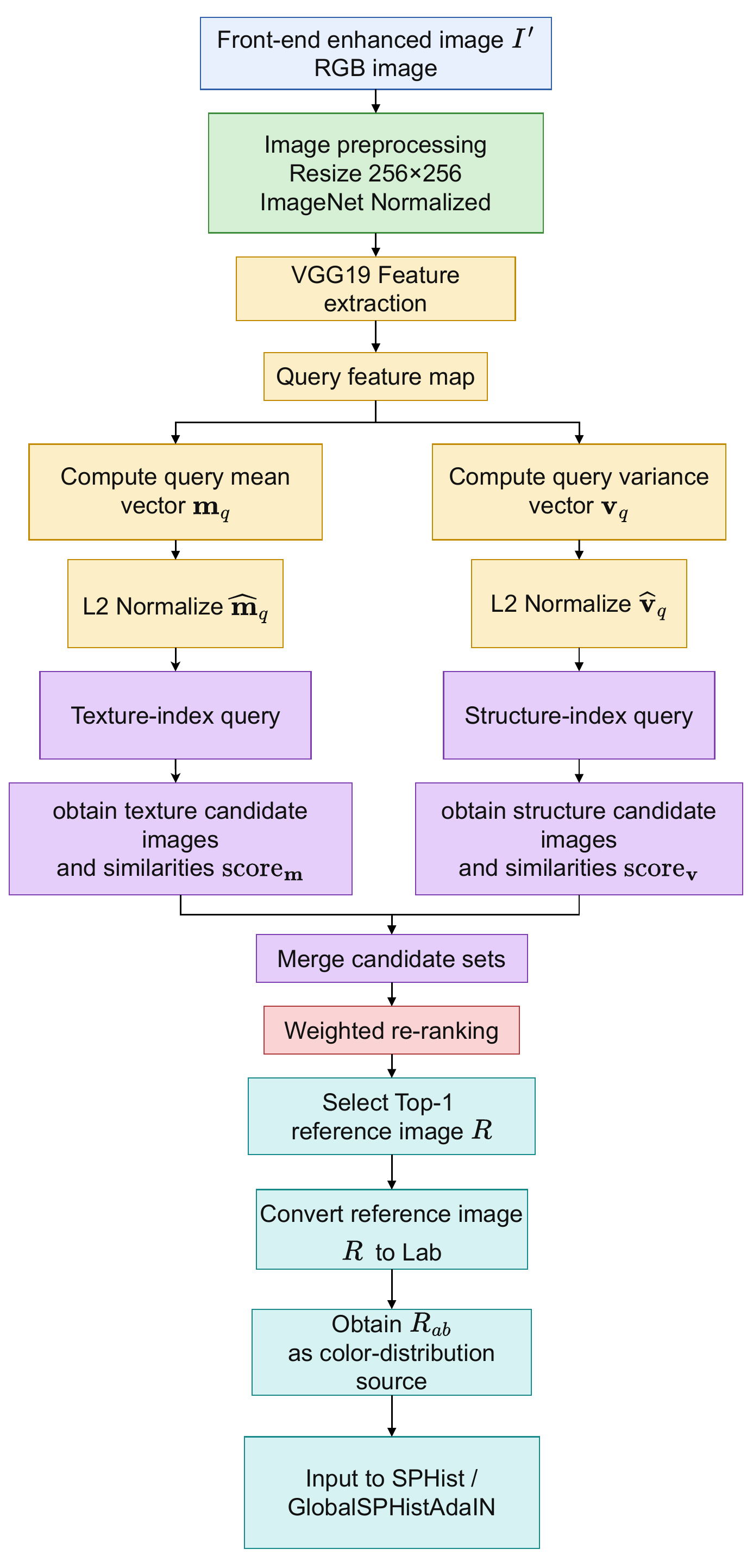}
\caption{Reference retrieval and weighted re-ranking. Each index returns candidates; scores are merged and re-ranked by $\alpha,\beta$ to select the Top-1 reference.}
\label{fig:rerank}
\end{figure}

\textit{Weighted re-ranking.} At query time, the enhanced RGB image passes through the same VGG19~\cite{ref12} pipeline to yield $\hat{\mathbf{m}}_q$ and $\hat{\mathbf{v}}_q$, which are searched against $\mathrm{Index}_{\mathbf{m}}$ and $\mathrm{Index}_{\mathbf{v}}$ respectively. For the $j$-th candidate, with texture score $\mathrm{score}_{\mathbf{m},j}$ and structure score $\mathrm{score}_{\mathbf{v},j}$, the combined similarity is
\begin{equation}
\mathrm{score}_j = \alpha \cdot \mathrm{score}_{\mathbf{m},j} + \beta \cdot \mathrm{score}_{\mathbf{v},j}
\end{equation}
with $\alpha$ and $\beta$ controlling the influence of texture and structure similarity; we set $\alpha = \beta = 0.5$ for equal weighting. Each index first returns $k_2 = 2 \times \text{Top-}k$ candidates; the two candidate sets are merged, and scores are accumulated per image, so an image appearing in both sets receives both contributions while an image in only one set receives only its own. Candidates are then sorted by $\mathrm{score}_j$ in descending order and the Top-$k$ are taken as references (Fig.~\ref{fig:rerank}). We set Top-$k = 1$ in training, validation, and inference. The selected reference $R$ is converted to Lab and its chromatic channels $R_{ab}$ become the input to color-vector extraction.

\subsection{Color-Restoration Module}

\begin{figure}[htbp]
\centering
\includegraphics[width=0.85\columnwidth]{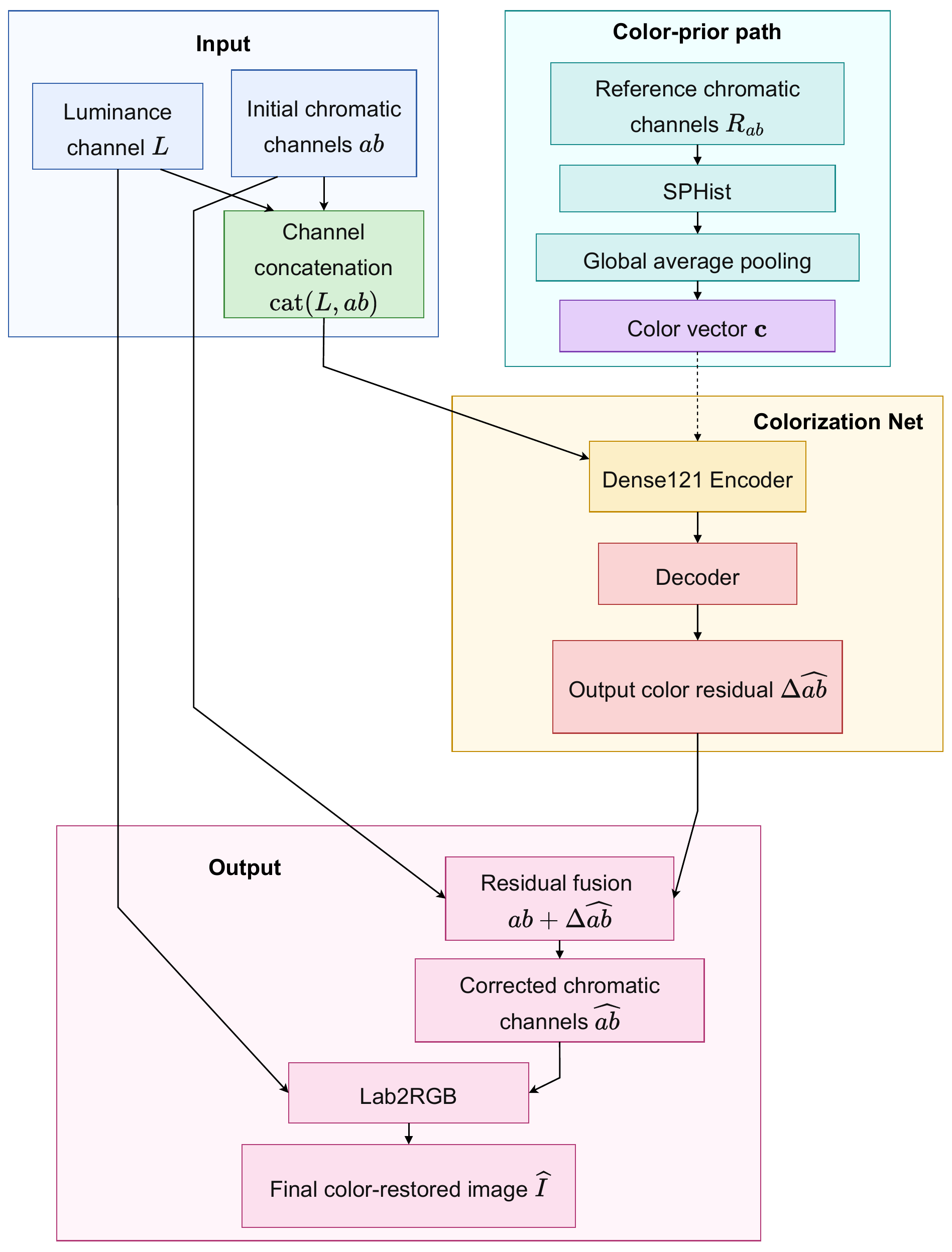}
\caption{Color-restoration module: SPHist color-vector extraction from the retrieved reference, GlobalSPHistAdaIN injection into encoder features, and residual color prediction.}
\label{fig:colormodule}
\end{figure}

We base the color-restoration module on the reference-based architecture of Pik-Fix~\cite{ref24}, but because a FAISS~\cite{ref11}-retrieved reference does not necessarily hold pixel-level correspondence with the input, the original spatial-alignment similarity sub-network is not applicable here. We therefore design GlobalSPHistAdaIN as a global color-modulation module that treats the reference as a source of global color prior and injects its color distribution without assuming spatial alignment (Fig.~\ref{fig:colormodule}).

\subsubsection{Colorization Network}
\begin{figure}[htbp]
\centering
\includegraphics[width=0.95\columnwidth]{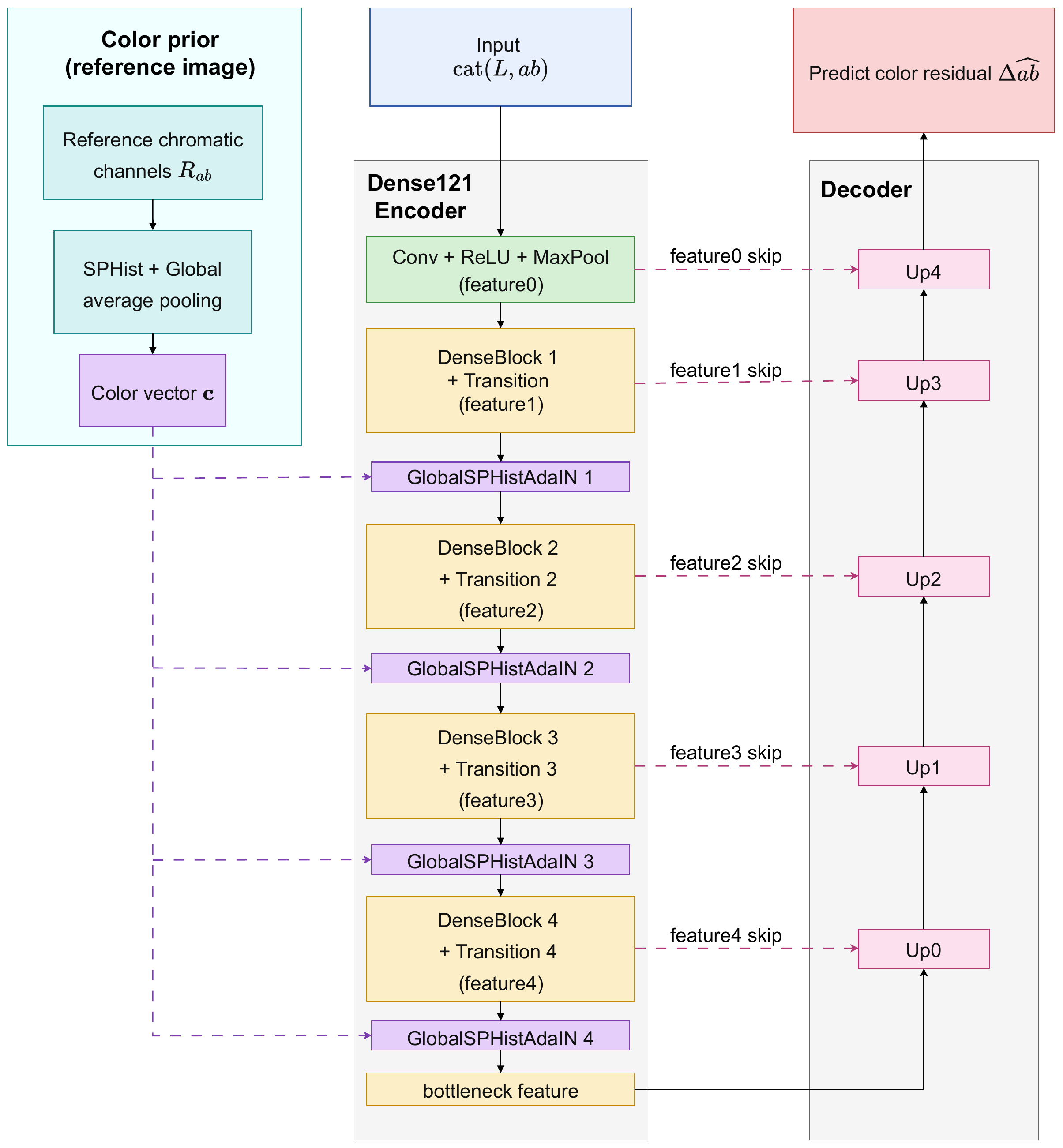}
\caption{Overall architecture of the color-restoration (colorization) network. The Dense-121 U-Net encoder injects the reference color prior through a GlobalSPHistAdaIN module after each transition layer; the decoder progressively restores resolution via skip connections and outputs the predicted color residual $\Delta\widehat{ab}$.}
\label{fig:encoder}
\end{figure}

The colorization network is a U-Net~\cite{ref26} (Fig.~\ref{fig:encoder}) whose encoder is based on DenseNet-121~\cite{ref32} with four dense blocks configured as $(6, 12, 24, 48)$ dense units and a growth rate of 32. Each dense block is followed by a transition layer that reduces spatial resolution and channel count---outputs of 128, 256, 512, and 1024 channels in turn---after which a GlobalSPHistAdaIN color-modulation module injects the reference's color prior into the deep encoder features; the modulated features then feed the subsequent encoding stage, propagating the color prior layer by layer. The decoder consists of five Up modules (up0--up4) that progressively restore resolution, each containing a residual dense block (RDB) and a double convolution, combined with the corresponding encoder scale through skip connections to preserve spatial detail.

 Within a dense block, the $\ell$-th layer takes the concatenation of all preceding feature maps as input,
\begin{equation}
x_\ell = H_\ell\big([x_0, x_1, \dots, x_{\ell-1}]\big)
\end{equation}
where $H_\ell(\cdot)$ is the $\ell$-th nonlinear transform and $[\cdot]$ denotes concatenation. The network input is the concatenation of the luminance channel and the front end's initial color channels,
\begin{equation}
\mathrm{cat}(L, ab) \in \mathbb{R}^{H \times W \times 3}
\end{equation}
where $L$ provides brightness and structure and $ab$ the front end's initial color estimate. The first convolution maps this 3-channel input to 64 initial channels; after the encoder, GlobalSPHistAdaIN modulation, and decoder reconstruction, a final convolution outputs a 2-channel color residual corresponding to the $a$ and $b$ channels.

\subsubsection{Spatial-Preserving Color Histogram (SPHist)}
\begin{figure}[htbp]
\centering
\includegraphics[width=0.55\columnwidth]{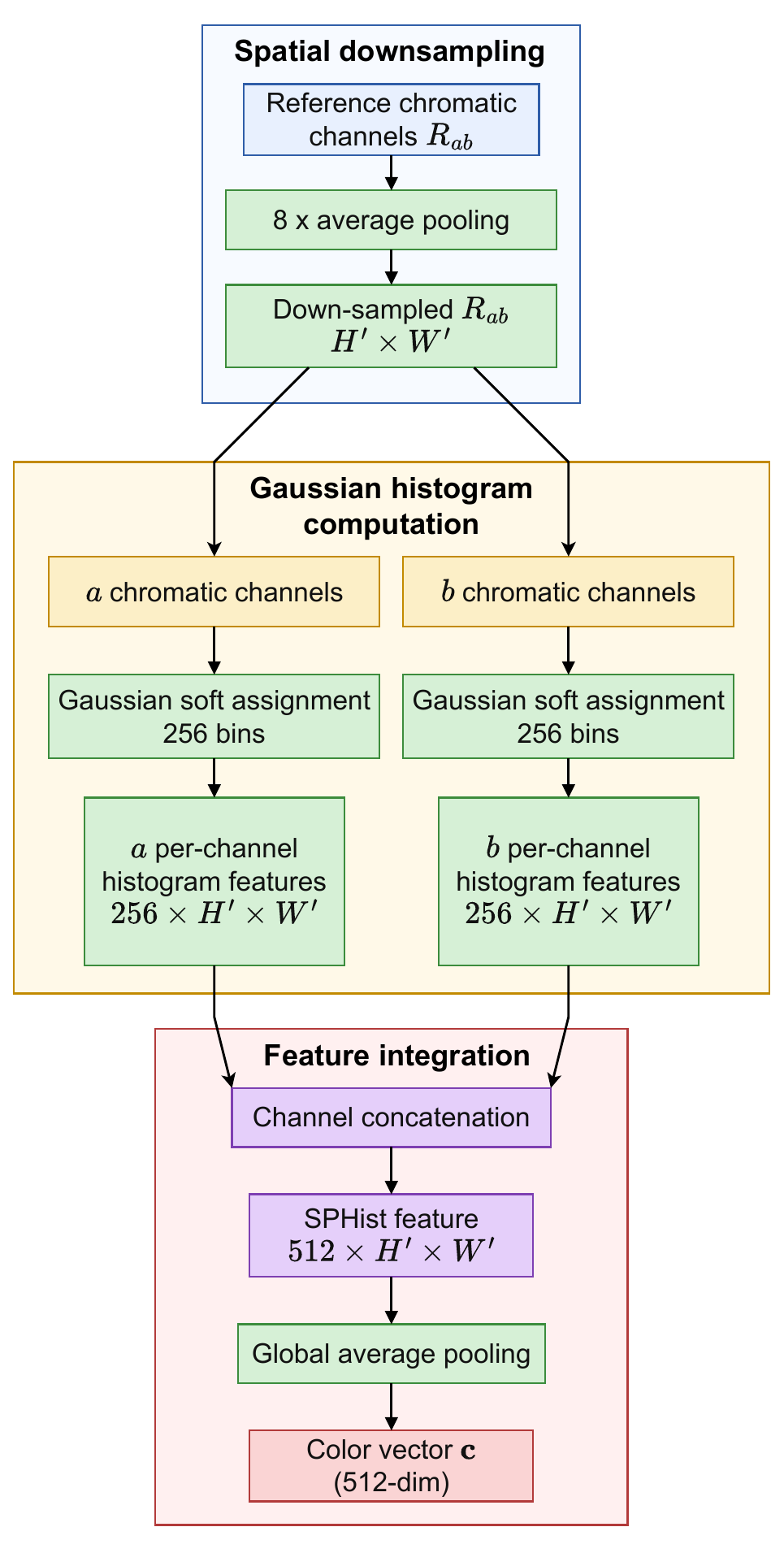}
\caption{Spatial-preserving color histogram (SPHist) computation via differentiable Gaussian soft assignment over the $a,b$ chromatic channels.}
\label{fig:sphist}
\end{figure}

To inject the reference's color information into the network effectively, we adopt the spatial-preserving color histogram introduced by Pik-Fix~\cite{ref24}. A conventional color histogram aggregates the whole image into a one-dimensional distribution and, in doing so, discards the spatial position of each pixel; SPHist instead preserves the per-pixel color distribution in a differentiable form suitable for training and inference (Fig.~\ref{fig:sphist}).

 SPHist approximates the $a$ and $b$ chromatic channels of $R_{ab}$ separately by Gaussian expansion. For the color value $D(i, j)$ at spatial position $(i, j)$, its response falling into the $k$-th bin is
\begin{equation}
h(i, j, k) = \frac{\exp\!\left(-\dfrac{(D(i, j) - u_k)^2}{2\sigma^2}\right)}{\sum_{k=1}^{K} \exp\!\left(-\dfrac{(D(i, j) - u_k)^2}{2\sigma^2}\right)}, \quad k = 1, 2, \dots, K
\end{equation}
where $u_k$ is the $k$-th fixed bin center, uniformly distributed over $[-1, 1]$; $K = 256$ is the number of bins per chromatic channel; and $\sigma = 0.01$ is the Gaussian kernel width. A small $\sigma$ makes each pixel's color value activate mainly its nearest bin centers. Since the $a$ and $b$ channels each yield a 256-bin spatial histogram, their concatenation forms an SPHist tensor $h_R \in \mathbb{R}^{512 \times H' \times W'}$. To reduce memory and computation, $R_{ab}$ is first $8\times$ average-pooled from $H \times W$ to $H' \times W'$ with $H' = H/8,\ W' = W/8$, after which global average pooling over the spatial dimensions produces a one-dimensional color vector $\mathbf{c} \in \mathbb{R}^{512}$:
\begin{equation}
\mathbf{c} = \frac{1}{H'W'} \sum_{h=1}^{H'} \sum_{w=1}^{W'} h_R(\cdot, h, w)
\end{equation}
The color vector $\mathbf{c}$ represents the reference's global color distribution over the $a, b$ chromatic space and serves as the conditioning input to GlobalSPHistAdaIN.

\subsubsection{Global Color-Modulation Module (GlobalSPHistAdaIN)}
\begin{figure}[htbp]
\centering
\includegraphics[width=\columnwidth]{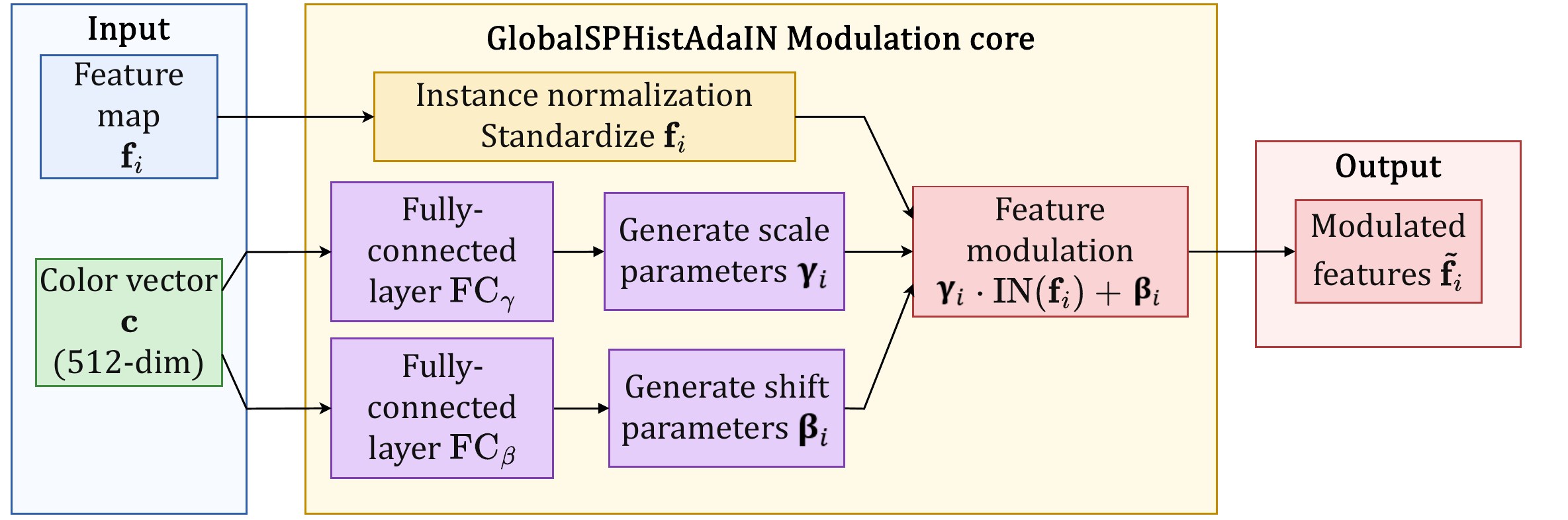}
\caption{GlobalSPHistAdaIN modulation flow: the 512-dimensional color vector generates per-channel scale and shift parameters applied to instance-normalized features.}
\label{fig:adain}
\end{figure}

The original Pik-Fix~\cite{ref24} computes a spatial correspondence between input and reference and warps the SPHist color information into alignment. In our cross-dataset setting, however, when a FAISS~\cite{ref11}-retrieved reference differs substantially in scene from the input, this correspondence is error-prone, and forcing a spatial warp can inject color from the wrong locations, producing color artifacts or unnatural shifts. We therefore design GlobalSPHistAdaIN to modulate intermediate features with the reference's global color distribution, improving stability and fault tolerance in cross-dataset inference (Fig.~\ref{fig:adain}).

 Its core mechanism is AdaIN~\cite{ref13}: given a feature map $\mathbf{f}$ with scale $\gamma$ and shift $\beta$,
\begin{equation}
\mathrm{AdaIN}(\mathbf{f}, \gamma, \beta) = \gamma \cdot \left(\frac{\mathbf{f} - \mu(\mathbf{f})}{\sigma(\mathbf{f})}\right) + \beta
\end{equation}
where $\mu(\mathbf{f})$ and $\sigma(\mathbf{f})$ are the per-channel mean and standard deviation under instance normalization. For the feature map $\mathbf{f}_i \in \mathbb{R}^{C_i \times H_i \times W_i}$ produced after each dense block's transition layer, the scale $\boldsymbol{\upgamma}_i$ and shift $\boldsymbol{\upbeta}_i$---vectors over the layer's $C_i$ channels---are generated from the color vector $\mathbf{c}$ by fully-connected layers:
\begin{equation}
\boldsymbol{\upgamma}_i = \mathbf{W}_{\gamma,i}\,\mathbf{c} + \mathbf{b}_{\gamma,i}, \qquad \boldsymbol{\upbeta}_i = \mathbf{W}_{\beta,i}\,\mathbf{c} + \mathbf{b}_{\beta,i}
\end{equation}
where $\mathbf{W}_{\gamma,i}, \mathbf{W}_{\beta,i} \in \mathbb{R}^{C_i \times 512}$ and $\mathbf{b}_{\gamma,i}, \mathbf{b}_{\beta,i} \in \mathbb{R}^{C_i}$. The modulated feature map is
\begin{equation}
\tilde{\mathbf{f}}_i = \boldsymbol{\upgamma}_i \cdot \mathrm{IN}(\mathbf{f}_i) + \boldsymbol{\upbeta}_i
\end{equation}
where $\mathrm{IN}(\cdot)$ is instance normalization and $\boldsymbol{\upgamma}_i, \boldsymbol{\upbeta}_i$ are broadcast to the spatial dimensions. Because GlobalSPHistAdaIN modulates only on the reference's global color statistics and does not rely on pixel-level alignment, it still provides a stable color-distribution prior when reference and input scenes differ, at lower cost than spatial-correspondence methods.

\subsubsection{Residual Learning}
Rather than predicting the full $ab$ channels, the network predicts a correction relative to the front end's $ab$:
\begin{equation}
\Delta\widehat{ab} = f_\theta\big(\mathrm{cat}(L, ab), \mathbf{c}\big)
\end{equation}
where $f_\theta(\cdot)$ is the network mapping with learnable parameters $\theta$. The residual target in training is
\begin{equation}
\Delta ab = ab_{\mathrm{GT}} - ab
\end{equation}
and the corrected color channels are
\begin{equation}
\widehat{ab} = ab + \Delta\widehat{ab}
\end{equation}
where $ab_{\mathrm{GT}}$ is the ground-truth color and $ab$ the front end's initial estimate. The residual formulation requires the network only to learn the difference between the front-end output and the true color; when the predicted residual approaches zero the result stays close to the front end's output, preserving the reasonable color the front end has already produced. The luminance channel $L$ and corrected color $\widehat{ab}$ are recombined and converted back to RGB,
\begin{equation}
\hat{I} = \mathrm{Lab2RGB}(L, \widehat{ab}), \quad \hat{I} \in \mathbb{R}^{H \times W \times 3}
\end{equation}

\subsection{Training Strategy}

Because the front end stays frozen, training targets only the second-stage color-restoration network. To eliminate any train--inference gap in the reference distribution, training uses the same FAISS~\cite{ref11} dual-index retrieval as inference.

\textit{Reference acquisition.} For each training sample, the front end's full enhanced image is the query; VGG19~\cite{ref12} extraction and FAISS~\cite{ref11} dual-index retrieval select the Top-1 reference from DIV2K~\cite{ref31}, which---after Lab conversion---provides $R_{ab}$ to the SPHist and GlobalSPHistAdaIN modules. Input and ground-truth images use identically-positioned random crops, while the reference remains the complete retrieved DIV2K~\cite{ref31} image resized to match. Using retrieval rather than synthetic or randomly chosen references at training time ensures the model learns color modulation under the same reference conditions it meets at inference.

\textit{Loss functions.} Although the network outputs a residual, the losses act not on $\Delta\widehat{ab}$ but on the final result after adding back the front-end color. The output is first clamped,
\begin{equation}
\widehat{ab} = \mathrm{clamp}(ab + \Delta\widehat{ab}, -1, 1)
\end{equation}
and losses are computed between $\widehat{ab}$ and $ab_{\mathrm{GT}}$. An L1 loss provides direct per-pixel chromatic supervision,
\begin{equation}
\mathcal{L}_{\mathrm{L1}} = \frac{1}{N} \sum_{i=1}^{N} \big| \widehat{ab}_i - ab_{\mathrm{GT},i} \big|
\end{equation}
where $N$ is the pixel count. A histogram loss based on the Earth Mover's Distance (EMD) constrains the color distribution: for the $a$ and $b$ channels we build 256-bin Gaussian histograms and approximate the EMD by the mean squared error of the cumulative distribution functions (CDF),
\begin{equation}
\mathcal{L}_{\mathrm{EMD}}(P, Q) = \frac{1}{K} \sum_{k=1}^{K} \big(\mathrm{CDF}_P(k) - \mathrm{CDF}_Q(k)\big)^2
\end{equation}
with $K = 256$. In practice the histogram loss combines the distance to the ground truth with the distance to the retrieved reference,
\begin{equation}
\mathcal{L}_{\mathrm{hist}}^{\mathrm{GT}} = \mathcal{L}_{\mathrm{EMD}}\big(H(\widehat{ab}), H(ab_{\mathrm{GT}})\big)
\end{equation}
\begin{equation}
\mathcal{L}_{\mathrm{hist}}^{\mathrm{ref}} = \mathcal{L}_{\mathrm{EMD}}\big(H(\widehat{ab}), H(R_{ab})\big)
\end{equation}
\begin{equation}
\mathcal{L}_{\mathrm{hist}} = \tfrac{1}{2}\big(\mathcal{L}_{\mathrm{hist}}^{\mathrm{GT}} + \mathcal{L}_{\mathrm{hist}}^{\mathrm{ref}}\big)
\end{equation}
where $H(\cdot)$ maps color channels to a Gaussian-histogram representation, $ab_{\mathrm{GT}}$ is the ground-truth color, and $R_{ab}$ is the retrieved reference color. An SSIM loss provides structural supervision by converting the prediction back to RGB,
\begin{equation}
\hat{I}_{\mathrm{RGB}} = \mathrm{Lab2RGB}(L, \widehat{ab})
\end{equation}
\begin{equation}
\mathcal{L}_{\mathrm{SSIM}} = 1 - \mathrm{SSIM}(\hat{I}_{\mathrm{RGB}}, I_{\mathrm{GT}})
\end{equation}
A perceptual loss supplements pixel-level supervision using deep features from a pretrained VGG16 network. Note that this VGG16 is used only for the perceptual loss and is distinct from the VGG19~\cite{ref12} used earlier for reference retrieval; the two serve different purposes. Rather than combining several layers, we use the feature map of a single fixed layer---the ReLU activation of the third convolutional block (relu3\_3) of VGG16---so that the perceptual loss is defined on one feature representation:
\begin{equation}
\mathcal{L}_{\mathrm{perc}} = \frac{1}{C\,H\,W} \big\lVert \phi(\hat{I}_{\mathrm{RGB}}) - \phi(I_{\mathrm{GT}}) \big\rVert_2^2
\end{equation}
where $\phi(\cdot)$ denotes the relu3\_3 feature map of VGG16 and $C$, $H$, $W$ are its channel, height, and width. The total loss is
\begin{equation}
\mathcal{L}_{\mathrm{total}} = \lambda_{\mathrm{L1}}\mathcal{L}_{\mathrm{L1}} + \lambda_{\mathrm{hist}}\mathcal{L}_{\mathrm{hist}} + \lambda_{\mathrm{SSIM}}\mathcal{L}_{\mathrm{SSIM}} + \lambda_{\mathrm{perc}}\mathcal{L}_{\mathrm{perc}}
\end{equation}

\textit{Datasets and setup.} We train and validate on LOLv1~\cite{ref3}, LOLv2-Real~\cite{ref33}, and LOLv2-Synthetic~\cite{ref33} independently. Before training, low-light images are enhanced by the Stage-1 front end, whose output feeds the color-restoration network with the corresponding normally-exposed image as ground truth. Training applies identically-positioned $256 \times 256$ random crops and random horizontal flips; images are converted to Lab with $L$ and $ab$ normalized to $[-1, 1]$. Validation and inference use full images, padded to multiples of 32 by reflect padding and cropped back afterward. We use Adam with a linear warmup followed by cosine-annealing decay, and train separate color-restoration networks for LLFormer~\cite{ref4}, FLIGHTNet~\cite{ref7}, and IAT~\cite{ref8} to test adaptability across front ends.

\section{Experiments}

\subsection{Experimental Setup}

\subsubsection{Datasets}
We evaluate on both paired and unpaired data. For paired evaluation we use three datasets. LOLv1~\cite{ref3} contains 485 training and 15 test pairs captured in real scenes at $400 \times 600$ resolution. LOLv2~\cite{ref33} extends LOLv1~\cite{ref3} into Real Captured (689 training, 100 test pairs) and Synthetic (900 training, 100 test pairs, generated with a camera-noise model) subsets. For unpaired evaluation we use LIME~\cite{ref2} (10 images), NPE~\cite{ref34} (84 images), MEF~\cite{ref35} (17 images), VV~\cite{ref36} (24 images), and DICM~\cite{ref37} (69 images), none of which provide normally-exposed ground truth.

\subsubsection{Metrics}
For paired data we report PSNR and SSIM for pixel- and structure-level fidelity, LPIPS~\cite{ref10} for perceptual similarity, and two color-specific metrics---CIEDE2000 ($\Delta E_{2000}$)~\cite{ref38} and the mean absolute error of the $ab$ channels ($\mathrm{MAE}_{ab}$). $\Delta E_{2000}$ is computed per pixel in Lab and averaged (scikit-image implementation), and $\mathrm{MAE}_{ab}$ is computed in the original Lab $ab$ range to preserve the native color-difference unit. For unpaired data we use NIQE~\cite{ref39}, a no-reference measure of natural-image statistics, treated as an auxiliary indicator.

\subsubsection{Implementation Details}
All experiments run on Windows 11 with Python 3.9 and PyTorch 2.8.0+cu128 on a single NVIDIA RTX 5070 (12~GB); NIQE~\cite{ref39} is evaluated in MATLAB and LPIPS~\cite{ref10} with the AlexNet backbone. The colorization network uses a Dense-121 U-Net encoder with $\text{block\_config} = (6, 12, 24, 48)$---the first three blocks inherit DenseNet-121~\cite{ref32} settings to load ImageNet-pretrained weights while the fourth is randomly initialized---and decoder RDBs with $\text{nDenseLayer} = [8, 12, 6, 4]$, growth rate 32, and dropout 0.3. GlobalSPHistAdaIN uses 256 fixed bin centers per $ab$ channel, forming the 512-dimensional color vector $\mathbf{c}$. Training uses Adam ($\beta_1 = 0.9$, $\beta_2 = 0.999$, $\varepsilon = 10^{-8}$, weight decay $10^{-5}$) with an initial learning rate of $10^{-4}$, a 5-epoch linear warmup, and cosine annealing to $10^{-6}$ over 100 epochs, at batch size 4 with $256 \times 256$ crops. Retrieval uses VGG19~\cite{ref12} relu3\_4 features (256-dim), two L2-normalized IndexFlatIP indices, a knowledge base of 800 DIV2K~\cite{ref31} training images, balanced weights $\alpha = \beta = 0.5$, and Top-$k = 1$. The loss weights are $\lambda_{\mathrm{hist}} = 2.0$, $\lambda_{\mathrm{L1}} = 0.5$, $\lambda_{\mathrm{SSIM}} = 0.1$, and $\lambda_{\mathrm{perc}} = 0.02$.

\subsection{Quantitative Results}

We report the two-stage architecture---front end plus RAG~\cite{ref9} color-restoration module---against the corresponding front-end-only baseline.

\subsubsection{LOLv1}
Table~\ref{tab:lolv1} gives results on LOLv1~\cite{ref3}. Adding the module improves most metrics across all four front ends, with $\Delta E_{2000}$ and $\mathrm{MAE}_{ab}$ falling most clearly; the more pronounced the front end's original color bias, the more room the module has to correct. The PSNR gain is largest for IAT~\cite{ref8} (+0.58~dB), followed by CPGA-Net++~\cite{ref6} (+0.29~dB), FLIGHTNet~\cite{ref7} (+0.20~dB), and LLFormer~\cite{ref4} (+0.06~dB); for our primary front end CPGA-Net++~\cite{ref6}, the module also lowers $\Delta E_{2000}$ from 8.912 to 8.402 and $\mathrm{MAE}_{ab}$ from 5.265 to 4.712. FLIGHTNet~\cite{ref7} + Proposed attains the best overall result (PSNR 25.16~dB, $\Delta E_{2000}$ 6.903, $\mathrm{MAE}_{ab}$ 4.611).

\begin{table}[h]
\centering
\caption{Quantitative results on LOLv1~\cite{ref3} (15 test pairs). Best per front-end pair in \textbf{bold}.}
\label{tab:lolv1}
\small
\begin{tabular}{lccccc}
\toprule
Method & PSNR$\uparrow$ & SSIM$\uparrow$ & LPIPS$\downarrow$ & $\Delta E_{2000}\downarrow$ & $\mathrm{MAE}_{ab}\downarrow$ \\
\midrule
CPGA-Net++~\cite{ref6} & 22.24 & 0.835 & 0.136 & 8.912 & 5.265 \\
\quad + Proposed & \textbf{22.53} & \textbf{0.836} & \textbf{0.128} & \textbf{8.402} & \textbf{4.712} \\
LLFormer~\cite{ref4} & 23.65 & 0.816 & 0.169 & 7.938 & 4.877 \\
\quad + Proposed & \textbf{23.71} & \textbf{0.817} & \textbf{0.168} & \textbf{7.881} & \textbf{4.789} \\
FLIGHTNet~\cite{ref7} & 24.96 & 0.849 & 0.134 & 7.206 & 4.905 \\
\quad + Proposed & \textbf{25.16} & \textbf{0.851} & \textbf{0.132} & \textbf{6.903} & \textbf{4.611} \\
IAT~\cite{ref8} & 23.38 & 0.806 & 0.216 & 7.974 & 5.611 \\
\quad + Proposed & \textbf{23.96} & \textbf{0.810} & \textbf{0.200} & \textbf{7.211} & \textbf{4.773} \\
\bottomrule
\end{tabular}
\end{table}

\subsubsection{LOLv2-Real Captured}
Table~\ref{tab:lolv2real} reports LOLv2-Real~\cite{ref33}; LLFormer~\cite{ref4} and IAT~\cite{ref8} are omitted as their official implementations provide no LOLv2 pretrained weights. The gain is more pronounced with FLIGHTNet~\cite{ref7}: FLIGHTNet~\cite{ref7} + Proposed raises PSNR by 0.26~dB, lowers $\Delta E_{2000}$ from 9.864 to 9.207 ($-6.7\%$), and lowers $\mathrm{MAE}_{ab}$ from 4.581 to 3.810 ($-16.8\%$). Under CPGA-Net++~\cite{ref6} the gain is smaller and LPIPS rises slightly, likely because this front end already provides a stable initial color estimate on LOLv2-Real~\cite{ref33}, leaving little residual to correct; even so, $\Delta E_{2000}$ and $\mathrm{MAE}_{ab}$ still improve.

\begin{table}[h]
\centering
\caption{Quantitative results on LOLv2-Real Captured~\cite{ref33} (100 test pairs).}
\label{tab:lolv2real}
\small
\begin{tabular}{lccccc}
\toprule
Method & PSNR$\uparrow$ & SSIM$\uparrow$ & LPIPS$\downarrow$ & $\Delta E_{2000}\downarrow$ & $\mathrm{MAE}_{ab}\downarrow$ \\
\midrule
CPGA-Net++~\cite{ref6} & 21.29 & 0.851 & \textbf{0.162} & 9.419 & 3.445 \\
\quad + Proposed & \textbf{21.37} & 0.851 & 0.164 & \textbf{9.378} & \textbf{3.423} \\
FLIGHTNet~\cite{ref7} & 21.71 & 0.834 & 0.186 & 9.864 & 4.581 \\
\quad + Proposed & \textbf{21.97} & \textbf{0.835} & \textbf{0.176} & \textbf{9.207} & \textbf{3.810} \\
\bottomrule
\end{tabular}
\end{table}

\subsubsection{LOLv2-Synthetic}
LOLv2-Synthetic~\cite{ref33} is the setting in which the module gains least (Table~\ref{tab:lolv2syn}): under CPGA-Net++~\cite{ref6} PSNR and SSIM even dip slightly. Because the synthetic degradation is regular and its color distribution relates stably to the ground truth, the front end's initial $ab$ already sits close to the true color, leaving little residual; a reference retrieved from DIV2K~\cite{ref31} that does not match the synthetic distribution may induce unnecessary correction. The module is thus better suited to correcting the color bias typical of real low-light scenes.

\begin{table}[h]
\centering
\caption{Quantitative results on LOLv2-Synthetic~\cite{ref33} (100 test pairs).}
\label{tab:lolv2syn}
\small
\begin{tabular}{lccccc}
\toprule
Method & PSNR$\uparrow$ & SSIM$\uparrow$ & LPIPS$\downarrow$ & $\Delta E_{2000}\downarrow$ & $\mathrm{MAE}_{ab}\downarrow$ \\
\midrule
CPGA-Net++~\cite{ref6} & \textbf{24.31} & \textbf{0.920} & \textbf{0.064} & \textbf{6.693} & \textbf{2.813} \\
\quad + Proposed & 24.29 & 0.913 & 0.066 & 6.802 & 2.998 \\
FLIGHTNet~\cite{ref7} & 25.12 & \textbf{0.928} & 0.064 & 6.510 & 3.143 \\
\quad + Proposed & \textbf{25.13} & 0.924 & 0.064 & \textbf{6.500} & \textbf{3.101} \\
\bottomrule
\end{tabular}
\end{table}

\subsubsection{Unpaired NIQE}
Table~\ref{tab:niqe} reports NIQE~\cite{ref39} for CPGA-Net++~\cite{ref6} on the five unpaired datasets. Adding the module raises NIQE~\cite{ref39} slightly on every set (mean 2.735$\rightarrow$2.834, +0.099). Since NIQE~\cite{ref39} measures conformity to natural statistics rather than color accuracy, we read this as a trade-off between color correction and naturalness, to be judged with the visual results and paired-data color metrics.

\begin{table}[h]
\centering
\caption{NIQE~\cite{ref39} on unpaired datasets (CPGA-Net++~\cite{ref6} front end); lower is better.}
\label{tab:niqe}
\small
\begin{tabular}{lccc}
\toprule
Dataset & CPGA-Net++~\cite{ref6} & + Proposed & $\Delta$ \\
\midrule
LIME~\cite{ref2} & 2.865 & 3.448 & +0.583 \\
NPE~\cite{ref34} & 3.043 & 3.084 & +0.041 \\
MEF~\cite{ref35} & 3.383 & 3.910 & +0.527 \\
VV~\cite{ref36} & 1.959 & 2.111 & +0.152 \\
DICM~\cite{ref37} & 2.524 & 2.618 & +0.094 \\
\midrule
Average & 2.735 & 2.834 & +0.099 \\
\bottomrule
\end{tabular}
\end{table}

\subsection{Ablation Studies}

All ablations use the LOLv1~\cite{ref3} test set with CPGA-Net++~\cite{ref6} as the front end and extend to LOLv2~\cite{ref33} only for cross-dataset generalization.

\subsubsection{Retrieval Strategy}
We compare the front end alone, the module with a random reference, and the module with a FAISS~\cite{ref11}-retrieved reference (Table~\ref{tab:abl_retrieval}). Even a random reference improves on the front end (PSNR +0.24~dB, $\Delta E_{2000}$ $-3.4\%$, $\mathrm{MAE}_{ab}$ $-8.7\%$), showing the network can use an external image's overall color distribution. FAISS~\cite{ref11} retrieval then pushes every metric further. The network and the retriever thus contribute at different levels: the former performs correction, the latter improves alignment by selecting a more suitable reference.

\begin{table}[h]
\centering
\caption{Retrieval-strategy ablation on LOLv1~\cite{ref3}.}
\label{tab:abl_retrieval}
\small
\begin{tabular}{lccccc}
\toprule
Method & PSNR$\uparrow$ & SSIM$\uparrow$ & LPIPS$\downarrow$ & $\Delta E_{2000}\downarrow$ & $\mathrm{MAE}_{ab}\downarrow$ \\
\midrule
CPGA-Net++~\cite{ref6} (no module) & 22.24 & 0.835 & 0.136 & 8.912 & 5.265 \\
RAG (random ref) & 22.48 & 0.833 & 0.131 & 8.608 & 4.806 \\
RAG (FAISS ref) & \textbf{22.53} & \textbf{0.836} & \textbf{0.128} & \textbf{8.402} & \textbf{4.712} \\
\bottomrule
\end{tabular}
\end{table}

\subsubsection{Residual vs. Direct Prediction}
We compare predicting a residual against directly predicting the full $ab$ across the three paired datasets (Table~\ref{tab:abl_residual}). On LOLv1~\cite{ref3}, direct prediction is marginally better in absolute numbers (underlined), but cross-dataset testing exposes a fundamental difference: direct prediction collapses to 19.68~dB on LOLv2-Real~\cite{ref33} (1.69~dB below residual) and 19.77~dB on LOLv2-Synthetic~\cite{ref33}. Residual learning---needing only a correction over the initial $ab$---retains the front end's estimate on unseen distributions and generalizes far more reliably.

\begin{table}[h]
\centering
\caption{Residual vs. direct prediction across paired datasets (CPGA-Net++~\cite{ref6} front end). The better value between Residual and Direct is \underline{underlined}; the Baseline is listed for reference only and is not part of this comparison.}
\label{tab:abl_residual}
\small
\begin{tabular}{llccccc}
\toprule
Dataset & Method & PSNR$\uparrow$ & SSIM$\uparrow$ & LPIPS$\downarrow$ & $\Delta E_{2000}\downarrow$ & $\mathrm{MAE}_{ab}\downarrow$ \\
\midrule
\multirow{3}{*}{LOLv1~\cite{ref3}} & Baseline & 22.24 & 0.835 & 0.136 & 8.912 & 5.265 \\
& + Residual & 22.53 & \underline{0.836} & \underline{0.128} & 8.402 & 4.712 \\
& + Direct & \underline{22.58} & 0.835 & \underline{0.128} & \underline{8.298} & \underline{4.630} \\
\midrule
\multirow{3}{*}{LOLv2-Real~\cite{ref33}} & Baseline & 21.29 & 0.851 & 0.162 & 9.419 & 3.445 \\
& + Residual & \underline{21.37} & \underline{0.851} & \underline{0.164} & \underline{9.378} & \underline{3.423} \\
& + Direct & 19.68 & 0.815 & 0.199 & 10.90 & 6.448 \\
\midrule
\multirow{3}{*}{LOLv2-Syn~\cite{ref33}} & Baseline & 24.31 & 0.920 & 0.064 & 6.693 & 2.813 \\
& + Residual & \underline{24.29} & \underline{0.913} & \underline{0.066} & \underline{6.802} & \underline{2.998} \\
& + Direct & 19.77 & 0.821 & 0.144 & 9.334 & 8.305 \\
\bottomrule
\end{tabular}
\end{table}

\subsubsection{GlobalSPHistAdaIN}
Removing GlobalSPHistAdaIN leaves the color network worse on every metric than the plain CPGA-Net++~\cite{ref6} baseline (Table~\ref{tab:abl_adain}). Absent an effective external color-prior injection, an added color network does not reliably improve the front-end output and can even disturb the initial estimate through its nonlinear transforms. The full module surpasses the baseline on all metrics (PSNR +0.29~dB, $\Delta E_{2000}$ $-5.7\%$, $\mathrm{MAE}_{ab}$ $-10.5\%$), making GlobalSPHistAdaIN the key injection module.

\begin{table}[h]
\centering
\caption{GlobalSPHistAdaIN ablation on LOLv1~\cite{ref3}.}
\label{tab:abl_adain}
\small
\begin{tabular}{lccccc}
\toprule
Method & PSNR$\uparrow$ & SSIM$\uparrow$ & LPIPS$\downarrow$ & $\Delta E_{2000}\downarrow$ & $\mathrm{MAE}_{ab}\downarrow$ \\
\midrule
CPGA-Net++~\cite{ref6} & 22.24 & 0.835 & 0.136 & 8.912 & 5.265 \\
\quad + Proposed (w/o SPHistAdaIN) & 22.09 & 0.833 & 0.139 & 8.965 & 5.420 \\
\quad + Proposed (full) & \textbf{22.53} & \textbf{0.836} & \textbf{0.128} & \textbf{8.402} & \textbf{4.712} \\
\bottomrule
\end{tabular}
\end{table}

\subsubsection{Effectiveness of Retrieval}
A natural concern is whether the gains come merely from the extra parameters and training of a post-processing network. We train a control that removes only the retrieval path---keeping the same backbone, residual design, losses, and training setup, but dropping the FAISS~\cite{ref11} dual index and GlobalSPHistAdaIN---so the network must predict residuals from training-data statistics alone. (With no reference, the histogram loss reduces to its ground-truth term, doubled to preserve scale.) As Table~\ref{tab:abl_norag} shows, this retrieval-free post-processing is slightly worse than the baseline on all five metrics. Color restoration requires a per-image color prior, which cannot be stored statically in weights and must be supplied dynamically at inference.

\begin{table}[h]
\centering
\caption{Retrieval-path ablation on LOLv1~\cite{ref3}.}
\label{tab:abl_norag}
\small
\begin{tabular}{lccccc}
\toprule
Method & PSNR$\uparrow$ & SSIM$\uparrow$ & LPIPS$\downarrow$ & $\Delta E_{2000}\downarrow$ & $\mathrm{MAE}_{ab}\downarrow$ \\
\midrule
CPGA-Net++~\cite{ref6} & 22.24 & 0.835 & 0.136 & 8.912 & 5.265 \\
\quad + retrieval-free residual & 22.09 & 0.833 & 0.139 & 9.064 & 5.448 \\
\quad + Proposed & \textbf{22.53} & \textbf{0.836} & \textbf{0.128} & \textbf{8.402} & \textbf{4.712} \\
\bottomrule
\end{tabular}
\end{table}

\subsubsection{Knowledge-Base Interchangeability}
To check whether the network learned to \emph{use} retrieved statistics or merely \emph{memorized} the DIV2K~\cite{ref31} prior, we replace DIV2K~\cite{ref31} with Flickr2K~\cite{ref40} (2,650 unseen natural images) at inference, rebuilding the indices identically and keeping all weights unchanged (Table~\ref{tab:abl_kb}). With the unseen base, every metric except SSIM still clearly beats the baseline, indicating a base-agnostic mechanism. Flickr2K~\cite{ref40} trailing DIV2K~\cite{ref31} slightly is expected, since training built its indices on DIV2K~\cite{ref31}. In practice, adapting to a domain with distinctive color characteristics requires only rebuilding the index with domain images---no retraining of the color network and no change to the frozen front end.

\begin{table}[h]
\centering
\caption{Knowledge-base interchangeability on LOLv1~\cite{ref3}.}
\label{tab:abl_kb}
\small
\begin{tabular}{lccccc}
\toprule
Method & PSNR$\uparrow$ & SSIM$\uparrow$ & LPIPS$\downarrow$ & $\Delta E_{2000}\downarrow$ & $\mathrm{MAE}_{ab}\downarrow$ \\
\midrule
CPGA-Net++~\cite{ref6} & 22.24 & 0.835 & 0.136 & 8.912 & 5.265 \\
\quad + Proposed (Flickr2K~\cite{ref40}) & 22.34 & 0.834 & 0.130 & 8.632 & 4.895 \\
\quad + Proposed (DIV2K~\cite{ref31}) & \textbf{22.53} & \textbf{0.836} & \textbf{0.128} & \textbf{8.402} & \textbf{4.712} \\
\bottomrule
\end{tabular}
\end{table}

\subsubsection{Retrieval Features and Weights}
Table~\ref{tab:abl_feat} evaluates retrieval strategies. All the VGG19~\cite{ref12} rows use the same features but differ in how those features are turned into an index: we describe each image by the global mean and variance of its VGG19~\cite{ref12} feature map and index texture and structure separately, then combine them with weights $\alpha$ (texture, mean) and $\beta$ (structure, variance). Among five such weightings, texture-dominant ($\alpha{=}1,\beta{=}0$) beats structure-dominant ($\alpha{=}0,\beta{=}1$, worst on all metrics), and balanced weighting ($\alpha{=}\beta{=}0.5$) is best or tied-best on most metrics. Replacing this VGG19~\cite{ref12} mean/variance index with a CLIP~\cite{ref14} feature index (ViT-B/32 or ViT-B/16) is worse on all metrics. The last two rows are cross-feature re-ranking schemes, in which one feature (e.g.\ VGG19~\cite{ref12}) first retrieves a candidate pool and the other feature (e.g.\ CLIP~\cite{ref14}) then re-ranks it; both orders also underperform the balanced VGG19~\cite{ref12} index. Color restoration therefore has different retrieval needs from generic semantic search: the mean and variance of VGG19~\cite{ref12} intermediate features better preserve texture, local structure, and low-level color statistics, making them the more appropriate similarity measure than high-level semantic features.

\begin{table}[h]
\centering
\caption{Retrieval-strategy comparison on LOLv1~\cite{ref3} (CPGA-Net++~\cite{ref6} front end).}
\label{tab:abl_feat}
\small
\begin{tabular}{lccccc}
\toprule
Retrieval & PSNR$\uparrow$ & SSIM$\uparrow$ & LPIPS$\downarrow$ & $\Delta E_{2000}\downarrow$ & $\mathrm{MAE}_{ab}\downarrow$ \\
\midrule
VGG19 $\alpha{=}0.5,\beta{=}0.5$ & \textbf{22.53} & 0.836 & \textbf{0.128} & \textbf{8.402} & 4.712 \\
VGG19 $\alpha{=}1.0,\beta{=}0.0$ & 22.50 & 0.835 & 0.129 & 8.412 & 4.754 \\
VGG19 $\alpha{=}0.7,\beta{=}0.3$ & \textbf{22.53} & 0.836 & \textbf{0.128} & \textbf{8.402} & 4.718 \\
VGG19 $\alpha{=}0.3,\beta{=}0.7$ & 22.49 & 0.836 & \textbf{0.128} & 8.416 & \textbf{4.668} \\
VGG19 $\alpha{=}0.0,\beta{=}1.0$ & 22.34 & 0.835 & 0.131 & 8.616 & 4.840 \\
CLIP ViT-B/32 & 22.47 & 0.836 & 0.129 & 8.549 & 4.727 \\
CLIP ViT-B/16 & 22.49 & 0.835 & 0.129 & 8.514 & 4.729 \\
VGG19 $\rightarrow$ CLIP rerank & 22.42 & 0.834 & 0.130 & 8.558 & 4.736 \\
CLIP $\rightarrow$ VGG19 rerank & 22.40 & 0.834 & 0.131 & 8.563 & 4.888 \\
\bottomrule
\end{tabular}
\end{table}

\subsection{Visual Analysis}

\begin{figure}[htbp]
\centering
\includegraphics[width=\columnwidth]{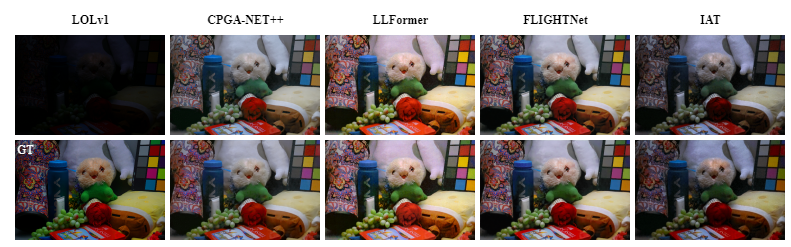}
\caption{Visual comparison on LOLv1. Top: low-light input and the four front-end outputs; bottom: ground truth and each front end paired with the proposed module.}
\label{fig:vis_lolv1}
\end{figure}

\textbf{Main comparison.} Figure~\ref{fig:vis_lolv1} shows LOLv1~\cite{ref3} results. All four front ends lift brightness to a recognizable level but retain some color bias---an overall warm cast, an under-saturated central green toy shifted toward yellow-green. After our module, the green toy's hue returns toward grass green with higher saturation, the fruit and other objects read more naturally, and the warm cast is suppressed; on the color chart, neutral-gray patches show reduced cast, bringing white balance closer to the ground truth. The trend holds across all four front ends.

\begin{figure}[H]
\centering
\includegraphics[width=\columnwidth]{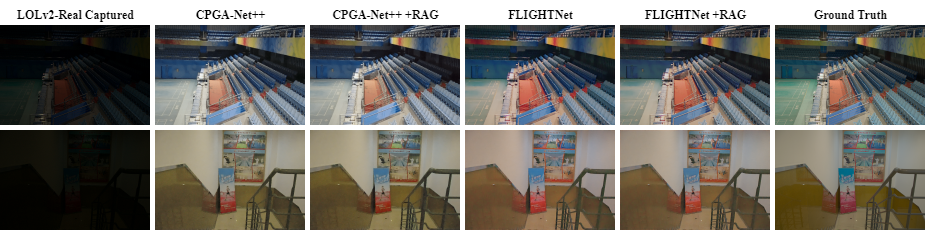}
\caption{Visual comparison on LOLv2-Real Captured across CPGA-Net++ and FLIGHTNet front ends with and without the proposed module.}
\label{fig:vis_lolv2real}
\end{figure}

\begin{figure}[H]
\centering
\includegraphics[width=\columnwidth]{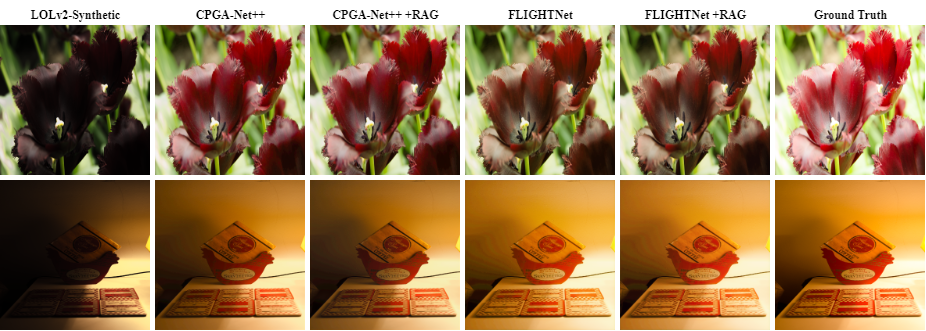}
\caption{Visual comparison on LOLv2-Synthetic; improvements are smaller due to the mismatch between synthetic color statistics and the natural-image knowledge base.}
\label{fig:vis_lolv2syn}
\end{figure}

\textbf{LOLv2 comparison.} Figures~\ref{fig:vis_lolv2real} and~\ref{fig:vis_lolv2syn} show LOLv2-Real~\cite{ref33} and LOLv2-Synthetic~\cite{ref33}. On real captures the module corrects front-end bias well---cool-blue indoor tones become natural warm white; cool-gray stairwell walls shift toward beige with improved saturation. On synthetic data the improvement is smaller and sometimes reversed, following from the reference source: retrieval from 800 DIV2K~\cite{ref31} natural images may not supply a prior consistent with algorithmically synthesized data.

\begin{figure}[H]
\centering
\includegraphics[width=0.75\columnwidth]{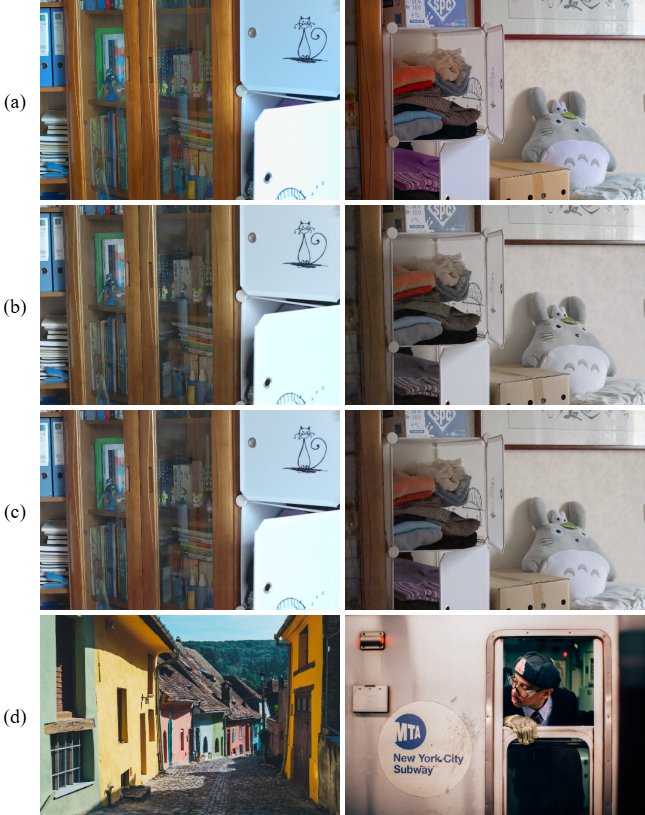}
\caption{Retrieval--restoration correspondence. Left: bookshelf scene, where a warm-dominant reference guides recovery of warm-brown tones. Right: near-neutral storage scene, where a low-saturation reference yields a near-zero residual. (a) ground truth; (b) front-end output; (c) proposed result; (d) retrieved reference.}
\label{fig:vis_corr}
\end{figure}

\textbf{Retrieval--restoration correspondence.} Figure~\ref{fig:vis_corr} illustrates the mechanism. In a bookshelf scene, a washed-out gray-brown wooden frame is corrected back to warm brown, guided by a retrieved warm-dominant streetscape, while the white cabinet stays neutral---saturation rises without a new cast. In a near-neutral storage scene, the retrieved low-saturation reference yields a near-zero residual and little change. The improvement thus depends on the residual color error left by the front end, not the module's intrinsic strength---consistent with the limited gains on LOLv2-Synthetic~\cite{ref33}.

\subsection{Efficiency and Complexity}

The system totals about 85.74~M parameters, of which the frozen CPGA-Net++~\cite{ref6} (0.06~M) and the VGG19~\cite{ref12} retrieval extractor (2.33~M) do not train; the trainable parameters are almost entirely in the colorization network (83.35~M), and the FAISS~\cite{ref11} index (1.65~MB) is a non-parametric vector store. Within the colorization network, the decoder Up modules dominate (65.0~M, 78.0\%) while the four GlobalSPHistAdaIN modules together are just 1.97~M (2.4\%). At $400 \times 600$, the colorization network needs about 853.06 GFLOPs against 28.94 GFLOPs for CPGA-Net++~\cite{ref6} and 29.34 GFLOPs for VGG19~\cite{ref12}, confirming that the main load is the second-stage network, not retrieval

Retrieval scales well: growing the base from 100 to 800 images increases the index linearly from 0.206 to 1.648~MB, yet end-to-end retrieval time stays near 6.6~ms, with FAISS~\cite{ref11} search at just 0.23~ms for 800 images. Over the full pipeline, mean per-image time is about 211.99~ms: colorization-network inference is the largest share (83.47~ms, 39.4\%), reference-image reading is the main non-compute cost (61.43~ms, 29.0\%, from loading high-resolution DIV2K~\cite{ref31} references off disk), and RAG~\cite{ref9} retrieval is only 7.40~ms (3.5\%). While not yet real-time, the compute is concentrated in the colorization network, and lightweighting that network and streamlining the I/O pipeline are the natural targets for future optimization.

\section{Conclusion}

We addressed the stalled progress of color restoration in low-light image enhancement with a framework built on retrieval-augmented generation. Our central observation is that although current LLIE models keep improving on luminance and structural metrics, gains in color accuracy lag well behind, and outputs still show systematic color shifts---indicating that color restoration is a relatively independent and under-addressed sub-problem. We therefore decoupled color restoration from the front-end LLIE model and realized it as a general-purpose post-processing module that retrieves a similar reference from an external color knowledge base and injects its color distribution to correct the front end's residual bias.

The design combines three elements. A dual-index FAISS~\cite{ref11} retriever over intermediate VGG19~\cite{ref12} features measures textural similarity through a global-mean index and structural similarity through a variance--covariance index, selecting the most suitable reference from a high-resolution DIV2K~\cite{ref31} base by weighted re-ranking. GlobalSPHistAdaIN injects color by reducing the reference's SPHist to a 512-dimensional color vector via global average pooling and modulating each layer of the color-restoration network through AdaIN~\cite{ref13}; removing spatial correspondence resolves the cross-dataset color-misinjection problem. A color-residual formulation trains the network to predict the difference between the front-end output and the ground truth rather than the full $ab$, so the network concentrates on correcting existing bias, improving training stability and cross-dataset generalization.

Systematic evaluation on three paired datasets bears this out. With CPGA-Net++~\cite{ref6} as the front end, the module lifts PSNR from 22.24 to 22.53~dB on LOLv1~\cite{ref3} while improving $\Delta E_{2000}$ from 8.91 to 8.40 and $\mathrm{MAE}_{ab}$ from 5.27 to 4.71, with color-specific metrics improving markedly more than brightness metrics. Swapping the front end for LLFormer~\cite{ref4}, FLIGHTNet~\cite{ref7}, or IAT~\cite{ref8} yields stable color-metric gains and reveals a clear trend---the weaker the front end's brightness enhancement, the larger the module's gain (PSNR +0.58~dB for IAT~\cite{ref8}, +0.20~dB for FLIGHTNet~\cite{ref7}, +0.06~dB for LLFormer~\cite{ref4}). Ablations separate each design choice: precise FAISS~\cite{ref11} retrieval improves on random references, and both beat the no-RAG~\cite{ref9} baseline; residual learning differs only marginally on LOLv1~\cite{ref3} but shows a decisive advantage (over 1.7~dB) on the more complex LOLv2~\cite{ref33} scenes; removing GlobalSPHistAdaIN degrades the network below the front-end baseline; and the VGG19~\cite{ref12} dual index beats CLIP~\cite{ref14} variants on all metrics, evidence that color restoration relies on textural and structural rather than semantic similarity. Replacing the training-time DIV2K~\cite{ref31} base with an unseen Flickr2K~\cite{ref40} base retains most of the gain, confirming that the network exploits retrieved color statistics rather than memorizing a fixed prior.

Several limitations remain. First, the module's effect depends heavily on the color-statistics match between input and reference base; with DIV2K~\cite{ref31} natural images it supplies effective priors on real captures but may misalign on synthetic data such as LOLv2-Synthetic~\cite{ref33}. Second, DIV2K~\cite{ref31} skews toward landscapes and architecture, offering limited coverage of professional scenes such as industrial inspection, medical imaging, or nighttime urban surveillance. Third, on synthetic data the residual space is minimal and the gain inherently small (PSNR change under 0.05~dB), suggesting such sets are ill-suited as the primary benchmark for a color-restoration module. Fourth, the module uses a Top-1 reference, leaving it dependent on a single retrieval's quality. Finally, the unpaired-data NIQE~\cite{ref39} results show a small naturalness decline, reflecting a trade-off in which correcting hue and chroma may perturb local statistics---so quality should be judged across all metrics together rather than by any single one.

These limitations point to four directions. The knowledge base can be expanded and made dynamic---switching the base by inference scene---corresponding to the Modular RAG~\cite{ref9} idea of composable retrievers. Multi-reference fusion and dynamic re-ranking offer another path: Top-$k$ SPHist averaging, attention-based fusion, and input-adaptive adjustment of the retrieval strategy. More broadly, the ``external knowledge base + retrieval + conditional generation'' philosophy can transfer to other restoration tasks governed by color or structural priors, where the differences lie chiefly in the base's content, the retrieval similarity measure, and the injection mechanism; building a unified visual-RAG~\cite{ref9} framework spanning several restoration tasks is a promising goal. In sum, this work introduces an independent color-restoration perspective to low-light image processing and provides a reproducible application case for RAG~\cite{ref9} in vision.

\section*{Declarations}
\textbf{Competing interests.} The authors declare no competing interests.\\
\textbf{Data availability.} All datasets used in this study (LOLv1/v2, LIME, NPE, MEF, VV, DICM, DIV2K, Flickr2K) are publicly available from their original publications.


\end{document}